\documentclass[11pt]{article}

\usepackage{amsmath,
            amssymb,
            mathtools,
            amsthm,
            bm,
            mathrsfs,
            stmaryrd,
            upgreek,
            textcomp,
            cancel,
            stackengine,
            eqlist,
            graphicx,
            xfrac
}

\usepackage[dvipsnames]{xcolor}
\usepackage[nosepfour,warning,np,debug,autolanguage]{numprint}

\usepackage{array,  % Already in the memoir class
            multirow,
            booktabs,  % Already in the memoir class
            float,
            rotating,
            caption,  % The memoir class loads ccaption
            afterpage,
            url%
}

\usepackage{subcaption}
\usepackage{threeparttablex}

\usepackage{enumerate} % Already in the memoir class
\usepackage{nth}

\usepackage{ifthen}
\usepackage{calc}
\usepackage{xspace}

\usepackage{acronym} % simple

\usepackage[T1]{fontenc}
\usepackage{hyperref}

\usepackage{geometry}
\usepackage[sort&compress]{natbib}

\theoremstyle{plain}

\theoremstyle{definition}
\newtheorem{remark}{Remark}

\usepackage{tikz}

\newcommand*\circinline[1]{%
  \begin{tikzpicture}
    \node[draw,circle,inner sep=0.5pt] {\scriptsize{#1}};
  \end{tikzpicture}}

\newcommand*\circd[1]{%
  \begin{tikzpicture}
    \node[draw,circle,inner sep=0.5pt]{\tiny{#1}};
  \end{tikzpicture}}

\newcommand{\scdot}{\ensuremath{\!\cdot\!}}
\newcommand{\Grad}[2][nonsense]{%
\ifthenelse{\equal{#1}{nonsense}}%
{\nabla#2}%
{\nabla_{#1}\,#2}%
}
\newcommand{\Div}[2][nonsense]{%
\ifthenelse{\equal{#1}{nonsense}}%
{\nabla\scdot#2}%
{\nabla_{#1}\scdot#2}%
}
\newcommand{\Lin}[2][nonsense]{%
\ifthenelse{\equal{#1}{nonsense}}%
{\ensuremath{\mathscr{L}(#2)}}%
{\ensuremath{\mathscr{L}(#1,#2)}}%
}

\renewcommand{\d}[2][nonsense]{%
\ifthenelse{\equal{#1}{nonsense}}%
{\ensuremath{\mathrm{d}#2}}%
{\ensuremath{\mathrm{d}^{#1}#2}}
}
\newcommand{\bv}[1]{\boldsymbol{\mathbf{#1}}}

\newcommand{\ts}[1]{\text{$\mathsf{#1}$}}
\newcommand{\transpose}[1]{#1^{\scriptscriptstyle\top}}
\newcommand{\inversetranspose}[1]{\ensuremath{#1^{\scriptscriptstyle-\top}}}
\newcommand{\order}[3][nonsense]{%
\ifthenelse{\equal{#1}{nonsense}}%
{\ensuremath{#2^{\scriptscriptstyle(#3)}}}%
{\ensuremath{#2^{\scriptscriptstyle#1(#3)}}}%
}

\newcommand{\colondot}{\mathbin{:}}

\newcommand{\s}{\ensuremath{\mathrm{s}}}
\newcommand{\f}{\ensuremath{\mathrm{f}}}
\newcommand{\m}{\ensuremath{\mathrm{m}}}
\newcommand{\flt}{\ensuremath{\mathrm{flt}}}

\newcommand{\va}{\ensuremath{\mathrm{va}}}
\newcommand{\ip}{\mathfrak{p}}
\newcommand{\g}{\ensuremath{\mathrm{g}}}

\newcommand{\comsol}{COMSOL Multiphysics\textsuperscript{\textregistered}\xspace}

\acrodef{CRF}[CRF]{constitutive response function}
\acrodef{SI}[SI]{International System of Units}
\acrodef{BVP}[BVP]{boundary value problem}
\acrodef{IBVP}[IBVP]{initial-boundary value problem}
\acrodef{FEM}[FEM]{finite element method}
\acrodef{FE}[FE]{finite element}
\acrodef{EIT}[EIT]{extension, inflation, and torsion}
\acrodef{ALE}[ALE]{arbitrary {L}agrangian--{E}ulerian}
\acrodef{FSI}[FSI]{fluid--structure interaction}
\acrodef{FPSI}[FPSI]{fluid--poroelastic structure interaction}

\acrodef{BDF}[BDF]{backward differentiation formula}
\acrodef{CNS}[CNS]{central nervous system}
\acrodef{CSF}[CSF]{cerebrospinal fluid}
\acrodef{IF}[IF]{interstitial fluid}
\acrodef{ECS}[ECS]{extracellular space}
\acrodef{SAS}[SAS]{subarachnoid space}
\acrodef{PVS}[PVS]{perivascular spaces}
\acrodef{TPE}[TPE]{theorem of power expended}
\acrodef{EOM}[EOM]{equation of motion}
\acrodefplural{EOM}[EsOM]{equations of motion}
\acrodef{HP}[HP]{Hamilton's principle}
\acrodef{BJS}[BJS]{Beavers--Joseph--Saffman}

\renewcommand{\circd}[1]{(\mathrm{#1})}
\renewcommand{\circinline}[1]{(\mathrm{#1})}

\usepackage{authblk}
\newcommand{\keywords}[1]{\par\medskip\noindent\textbf{Keywords:} #1}

\title{A Finite Element Model for Fluid--Poroelastic Structure Interaction with Dissipative Interface}
\author{Mohammad Jannesari}
\author{Francesco Costanzo\thanks{Corresponding author: fxc8@psu.edu}}
\affil{Center for Neural Engineering, University Park, PA 16802, USA\\
Engineering Science and Mechanics Department, University Park, PA 16802, USA}
\date{June 2026}

\begin{document}

\maketitle

\begin{abstract}
This paper extends our prior work on interface conditions between two poroelastic domains to the case when a poroelastic domain interacts with a pure
fluid. The porous medium is assumed to be deformable and modeled following mixture theory. When not in a mixture, each constituent of the porous domain is assumed to be incompressible. An arbitrary Lagrangian--Eulerian formulation is developed for both the fluid and porous domains.
The fluid velocity and pressure fields are allowed to be discontinuous across the pure fluid/porous domain interface. A globally continuous displacement field is defined to represent both the deformation of the solid skeleton of the poroelastic medium and the mesh motion in the pure fluid domain. As an interface condition, the continuity of the normal component of the filtration velocity is enforced via a Lagrange multiplier. The mixture momentum jump condition is implemented so as to transfer the traction from the fluid side to the porous domain.
A dissipative behavior is attributed to the interface, and the jump condition for the fluid momentum, derived from Hamilton's principle, incorporates this prescribed behavior. The model implementation is verified using the method of manufactured solutions, and a sample problem is solved to compare an inert interface with a fully dissipative one. The results show satisfactory convergence rates, while the proposed model successfully reproduces the analytical solution of a benchmark problem using conventional jump conditions. Furthermore, the findings highlight the significant influence of the selected interface model on the behavior of the system.
\end{abstract}

\section{Introduction}
The interaction of a free fluid and a deformable, fluid-saturated porous medium arises in many multiphysics problems, with application in biological tissue mechanics \cite{Kedarasetti2020Functional-hype,Kedarasetti2022Arterial-vasodi,Shim2021FiniteElementIm,Hou1989BoundaryConditi}, groundwater hydrology \cite{Anghileri2005FluidStructureI}, and energy storage systems \cite{Huang2026FluidStructureI}.
This interaction plays a significant role in the determination of the overall system response. A key and enduring challenge in the modeling of such multiphysics problems is the accurate representation of the \emph{transfer} of mass, momentum, and energy across deformable interfaces, where the interfacial constitutive behavior contributes to the system's response. The resulting problem inherits the mathematical and computational challenges of both classical \ac{FSI} and multiphase porous-media flow. Such problems are referred to as \ac{FPSI}.

The majority of the \ac{FPSI} literature models the porous region using Biot's theory of poroelasticity, in which the elastic deformation of the solid skeleton is coupled to a Darcy-type description of the pore fluid through a single effective pressure and a Biot--Willis coupling coefficient \cite{Avalos2024WeakAndStrongSo,Guo2024AFullyParalleli}.
Building on this formulation, a substantial body of work has addressed the coupling of Stokes or Navier--Stokes flow to the Biot system. The well-posedness of the coupled Stokes--Biot \cite{Avalos2025UniquenessOfWea} and Navier--Stokes--Biot \cite{Cesmelioglu2023AHybridizableDi} problems has been established under various formulations, and a range of monolithic \cite{Badia2009CouplingBiotAnd} and partitioned \cite{Bukac2015PartitioningStr} solution strategies have been developed and analyzed, including Lagrange-multiplier \cite{Castro2025WellPosednessOf,Wang2025ALagrangeMultip} and Robin--Robin \cite{Dalal2024ARobinRobinSpli} approaches.
While the Biot framework offers a compact and well-studied description of poroelastic behavior, it represents the pore fluid only through its volume-averaged velocity and a single pressure field, which limits its ability to resolve the individual dynamics of the solid and fluid constituents, their distinct stress states, and the momentum exchange between them. An alternative, more general description is provided by mixture theory \cite{Bowen1976Theory-of-Mixtu0,Bowen1980Incompressible-0}, in which the porous medium is treated as a superposition of interpenetrating solid and fluid phases, each satisfying its own balance laws, coupled through interphase momentum-exchange (drag) terms.
This framework, which has been widely developed for multiphase and multi-constituent porous media, enables the individual velocities and stresses of each phase to be described explicitly.
As a result, it provides a natural framework for formulating the interface conditions required when two poroelastic mixtures interact \cite{Costanzo2026PoroelasticFlow} or when biphasic models are used to represent biological tissues and engineered scaffolds \cite{Jannesari2026ModificationOfA}.

Various interface conditions have been proposed for the interface between contiguous porous domains \cite{Costanzo2026PoroelasticFlow,dellIsola2009Boundary-Condit-0}. However, the situation changes fundamentally when one side of the interface is a free fluid rather than a porous continuum, giving rise to challenges including, but not limited to, the appropriate interfacial pseudo-slip condition and the kinematic coupling at a deforming and permeable boundary that may itself evolve over time.
Moreover, the interface may behave as an inert membrane, neither storing nor dissipating energy, or it may exhibit its own constitutive behavior, thereby necessitating additional interface conditions beyond the jump conditions implied by the balance equations.

In the literature on coupling a free fluid with a poroelastic domain, interface conditions are usually grouped according to the physical quantity they constrain, regardless of whether the poroelastic domain is described by Biot's model or a mixture-theory formulation.
The first and most commonly used condition is continuity of the normal flux. This means that the normal component of the free-fluid velocity is equal to the normal component of the total flux leaving the poroelastic domain.
This condition follows directly from conservation of mass at the interface, assuming that no fluid is stored or generated there, and it is used in essentially the same form for different free-fluid and porous-medium models \cite{Discacciati2007RobinRobinDomai,Badia2009CouplingBiotAnd}.

The second condition deals with the normal stress and is more subtle in what it assumes. The classical formulation, introduced in Showalter's work on poroelastic filtration coupled with Stokes flow \cite{Showalter2005PoroelasticFilt}, states that the free fluid's total Cauchy stress is balanced by the total stress in the poroelastic medium.
A different version of this condition is obtained by using rigorous two-scale homogenization of the pore geometry instead of assuming it from the start. In this version, pressure is not required to be continuous across the interface. Instead, there is a pressure jump that depends on the permeability and the tangential strain rate of the free fluid at the interface. This condition was confirmed by comparing it with direct numerical simulations of the pore-scale problem \cite{Carraro2013PressureJumpInt}.

The third, and historically the most contested, condition governs the tangential direction. The original Beavers--Joseph condition \cite{Beavers1967BoundaryConditi} was proposed empirically from channel-flow experiments over a Darcy-type porous bed and relates the tangential shear stress in the free fluid to the difference between the tangential free-fluid velocity and the tangential filtration velocity, scaled by a dimensionless slip coefficient dependent on the pore microstructure; Saffman's subsequent simplification \cite{Saffman1971OnTheBoundaryCo} assumes the tangential filtration velocity is negligible relative to the free-fluid velocity and so restates the law using only the tangential free-fluid velocity, an approximation later extended to general tangential directions \cite{Jones1973LowReynoldsNumb}.

In the mixture-theory literature, interface conditions for fluid--poroelastic (or biphasic--biphasic) coupling are usually described using three main requirements. The first is continuity of pressure, meaning that the pore or interstitial fluid pressure is the same on both sides of the interface. This is the condition most commonly used in biphasic biomechanics. It was first introduced for the cartilage--synovial fluid interface \cite{Hou1989BoundaryConditi} and has been used in later biphasic contact and fluid--structure models \cite{Ateshian2010FiniteElementAl,Shim2021FiniteElementIm}. However, this condition is not always imposed. Variational mixture-theory derivations allow a pressure jump that is determined by an interfacial constitutive law \cite{dellIsola2009Boundary-Condit-0,Costanzo2026PoroelasticFlow}, while homogenization-based derivations on the free-fluid side also predict a permeability-dependent pressure jump instead of strict pressure continuity \cite{Carraro2013PressureJumpInt}.
The second requirement is continuity of the normal component of the filtration velocity, or equivalently, continuity of the normal fluid flux. This condition is applied more consistently than pressure continuity. It was included in the original biphasic interface conditions \cite{Hou1989BoundaryConditi}, in biphasic contact models that require continuity of the normal fluid flux across the interface \cite{Ateshian2010FiniteElementAl}, and in the hybrid biphasic mixture jump conditions \cite{Shim2021AHybridBiphasic}, where it is derived from the balance of mass and energy rather than assumed directly.
The third requirement describes the tangential behavior at the interface and is the least well established. Hou et al. \cite{Hou1989BoundaryConditi} introduced a ``pseudo-no-slip'' condition, which requires continuity of the tangential component of the relative fluid flux instead of the fluid velocity itself. They showed that this condition reduces to the classical Beavers--Joseph slip law for a viscous fluid flowing over a rigid porous medium. Other studies instead impose continuity of the tangential viscous traction, which requires the effective fluid viscosity to vary with the local fluid volume fraction across the interface.

In our previous work, we investigated poroelastic flow across a permeable interface within a Hamiltonian framework \cite{Costanzo2026PoroelasticFlow}. We derived and analyzed a general set of interface conditions for the coupling of two poroelastic domains, each governed by mixture theory.
In this paper, we extend our previous framework to \ac{FPSI}. One domain includes a free, single-phase incompressible viscous fluid described by the Navier--Stokes equations. The other domain consists of a poroelastic medium described by mixture theory.
This expansion does not come from merely adjusting the interface conditions for two poroelastic media. The free-fluid region has no solid phase or associated volume fraction, which fundamentally changes the interfacial coupling. Thus, we need to modify the jump conditions for the balance equations. In addition, a typical dissipative constitutive model is adopted for the interface to examine how the overall system response differs between an inert interface and a dissipative one.
Because \ac{FPSI} usually involves large interface deformations and considerable relative motion between the two domains, we formulate the problem within an \ac{ALE} framework \cite{Costanzo2017AnArbitraryLagr,Fernandez2009TheDerivationOf}. 
Taken together, these contributions extend previously established interface conditions to \ac{FPSI} within a unified ALE framework while also demonstrating a practical and validated implementation strategy. The resulting approach provides a versatile tool for analyzing problems in which interface motion plays a central role and establishes a solid foundation for future developments.

The rest of the paper is organized as follows. Section~\ref{sec: basic defintions and kinematics} presents the kinematics for the poroelastic and pure fluid domains.
Section~\ref{sec: balance of mass and momentum} introduces the balance equations for both domains, while the interface conditions that couple these two domains are presented in Section~\ref{sec: interface conditions}.
In Section~\ref{sec: constitutive response functions}, we introduce the constitutive response functions for each of the domains.
Section~\ref{sec: weak formulation} discusses the corresponding weak formulation, together with the functional spaces for the principal variables.
Section~\ref{sec: results} assesses the proposed approach through verification using manufactured solutions and validation against a benchmark problem.
The setup and results of a sample problem comparing different interface behaviors are presented in Section~\ref{sec: results sample problem}.
Finally, Section~\ref{sec: discussion and conclusion} summarizes the main findings and discusses their implications, along with potential directions for future research.

\section{Basic Definitions and Kinematics}
\label{sec: basic defintions and kinematics}

The set of real numbers is denoted by $\mathbb{R}$, with $\mathbb{R}^{+}$ and $\mathbb{R}^{+}_{0}$ representing its positive and non-negative subsets, respectively. Time is introduced as $t\in \mathbb{R}^{+}_{0}$ . $\mathcal{E}^{d}$ designates the d-dimensional Euclidean point space, while its translation vector space is represented by $\mathscr{V}$. For any given vector spaces $\mathscr{U}$ and $\mathscr{W}$, we designate the vector space of all linear operators with domain $\mathscr{U}$ and range in $\mathscr{W}$ by $\Lin[\mathscr{U}]{\mathscr{W}}$. The notation $\Lin{\mathscr{V}}$ serves as a concise abbreviation for $\Lin[\mathscr{V}]{\mathscr{V}}$.

The motion of the system and the related definitions are introduced in the following sections. The presentation begins with the porous-medium framework, which is formulated according to the mixture theory approach in \cite{Bowen1976Theory-of-Mixtu0,Bowen1980Incompressible-0}. Subsequently, the \ac{ALE} formulation for the fluid domain is described.
The \ac{ALE} description combines the benefits of both Lagrangian and Eulerian approaches. It allows the computational mesh to move independently of the material while tracking the evolution of the fluid--poroelastic interface.

Let $\Sigma$ denote a smooth codimension-1 submanifold of a d-dimensional domain $\mathcal{A}\subset\mathcal{E}^{d}$, and let $\bv{\nu}$ be a unit vector field defining its orientation. Consider a field $\phi(\bv{X})$, with $\bv{X}\in\mathcal{A}$, that is continuous everywhere except possibly across $\Sigma$. Then, for $\hat{\bv{X}}\in\Sigma$,
%+-
\begin{equation*}
\phi^{\pm} \coloneqq
\lim_{s\to0^{+}} \phi(\hat{\bv{X}}\pm s \bv{\nu}),\quad
\llbracket\phi\rrbracket \coloneqq \phi^{+} - \phi^{-},
\quad\text{and}\quad
\langle\phi\rangle \coloneqq \tfrac{1}{2}(\phi^{+}+\phi^{-}).
\end{equation*}
%+-

\subsection{Kinematics}

\begin{figure}[hbt]
\centering
\includegraphics[width=0.5\textwidth]{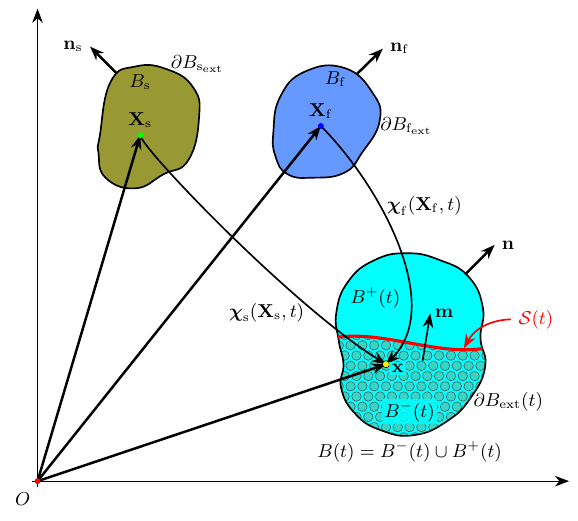}
\caption{The \ac{FPSI} problem is described by its current configuration $B(t)$, which comprises the current solid and fluid configurations. The configurations $B_{\s}$ and $B_{\f}$ denote the reference configurations of the solid and fluid domains, respectively. A point in $B_{\s}$ is indicated by $\bv{X}_{\s}$, a point in $B_{\f}$ is indicated by $\bv{X}_{\f}$, and $\bv{x}$ denotes points in $B(t)$. For $B_{\s}$, $B_{\f}$, and $B(t)$, the corresponding boundaries are $\partial B_{\s_{\mathrm{ext}}}$, $\partial B_{\f_{\mathrm{ext}}}$, and $\partial B_{\mathrm{ext}}(t)$. Each boundary carries an orientation given by an outward unit normal field, namely $\bv{n}_{\s}$, $\bv{n}_{\f}$, and $\bv{n}$, respectively. $\mathcal{S}(t)$ denotes the interface separating the pure fluid domain from the biphasic mixture, and it is oriented by $\bv{m}$.}
\label{fig: fpsi problem definition}
\end{figure}

We consider a mixture of solid and fluid interacting with an external pure fluid of the same kind. Both the mixture and the pure fluid domain, in their deformed state, define the current configuration, denoted by $B(t) \subset \mathcal{E}^{d}$ with $d=2,3$, whereas $B_{\s} \subset \mathcal{E}^{d}$ and $B_{\f} \subset \mathcal{E}^{d}$ denote the reference configurations of the solid and the fluid, respectively, as illustrated in Fig.~\ref{fig: fpsi problem definition}.
%+-
%+-
Points belonging to $B(t)$, $B_{\s}$, and $B_{\f}$ are respectively denoted by $\bv{x}$, $\bv{X}_{\s}$, and $\bv{X}_{\f}$.
The motions of the solid and fluid phases are defined as
%+-
\begin{equation}
\label{eq: Motions defs}
\bv{\chi}_{\s}(\bv{X}_{\s},t): B_{\s} \times \mathbb{R}_{0}^{+}\to B(t)
\quad\text{and}\quad
\bv{\chi}_{\f}(\bv{X}_{\f},t): B_{\f} \times \mathbb{R}_{0}^{+}\to B(t),
\end{equation}
%+-
such that 
$$\bv{x}=\bv{\chi}_{\s}(\bv{X}_{\s},t)=\bv{\chi}_{\f}(\bv{X}_{\f},t).$$

We introduce the following time-independent partition of $B_{\s}$:
%+-
\begin{equation}
\label{eq: label Bs partition}
B_{\s} = B_{\s}^{+} \cup B_{\s}^{-},\quad
\mathring{B}_{\s}^{+} \cap \mathring{B}_{\s}^{-} = \emptyset,\quad\text{and}\quad B_{\s}^{+} \cap B_{\s}^{-} = \mathcal{S}_{\s},
\end{equation}
%+-
where, for a set $A$, $\mathring{A}$ denotes the interior of $A$, and $\mathcal{S}_{\s}$ is assumed to be a regular nonintersecting surface. While we view $\bv{\chi}_{\s}$ as being globally continuous over $B_{\s}$, we will treat the physical properties of the solid phase, e.g., mass density distribution, as being nontrivial only over $B_{\s}^{-}$. The images of $B_{\s}^{+}$ and $B_{\s}^{-}$ under $\bv{\chi}_{\s}$ provide a corresponding partition of $B(t)$, namely $B^{+}(t)$ and $B^{-}(t)$, characterized by an interface $\mathcal{S}(t)$. The physical manifestation of the solid phase in the current configuration is therefore limited to the domain $B^{-}(t)$. By contrast, the fluid phase is understood to have nontrivial physical behavior in all of $B(t)$. Under this stipulation, we refer to $B^{+}(t)$ as the pure fluid domain, whereas we refer to $B^{-}(t)$ as the poroelastic domain. We note that $\mathcal{S}(t)$ can be permeable, in which case fluid particles are allowed to move across $\mathcal{S}(t)$.

\begin{remark}[Notation for domain-specific quantities]
Whenever necessary, the superscripts $+$ and $-$ are used to indicate to which domain a given function pertains: $+$ denotes association with the pure fluid, while $-$ denotes association with the porous medium.
\end{remark}

\begin{remark}[Meaning of $\bv{\chi}_{\s}$ over $B^{+}_{\s}$---\ac{ALE} framework]
The partition of $B_{\s}$ in Eq.~\eqref{eq: label Bs partition} allows us to give different meanings to $\bv{\chi}_{\s}$ over the partition. Letting $\phi\mid_{A}$ denote the restriction of a function $\phi$ to the subset $A$, we view $\bv{\chi}_{\s}\mid_{B^{-}_{\s}}$ as representing the physical motion of the solid phase. Instead, we view $\bv{\chi}_{\s}\mid_{B^{+}_{\s}}$ as providing an \ac{ALE} map suitable for building a convenient implementation of the physics of the fluid phase and of the jump conditions across the interface $\mathcal{S}(t)$. Therefore, as is typical in \ac{ALE} formulations, the equations governing  $\bv{\chi}_{\s}\mid_{B^{+}_{\s}}$ will be chosen by convenience and subject only to a global continuity requirement across $\mathcal{S}_{\s}$. In this view, $\bv{\chi}_{\s}\mid_{B^{+}_{\s}}$ can be considered a suitable continuous extension of the physical motion of the solid phase over the pure fluid domain. Having established the meaning of $\bv{\chi}_{\s}$, when it comes to the physics of fluid phase, we will use $\bv{\chi}_{\s}$ as a global \ac{ALE} map to pull back the physics of the fluid phase to the reference configuration $B_{\s}$. In our examples, we will choose $B_{\s}$ as a computational domain. Again, the physics of the solid is limited to a subset of this domain, whereas the physics of the fluid, under the action of the \ac{ALE} map $\bv{\chi}_{\s}$ (or, more specifically, its inverse) is supported over all of $B_{\s}$.
\end{remark}

\begin{remark}[Concerning the domain of functions]
To differentiate the original form of a function from the version resulting from its composition with domain mappings, the notation proposed by \cite{dellIsola2009Boundary-Condit-0} is adopted. For $a,b = \s,\f, \m$, consider smooth functions $\varphi: B_{a}\to\mathscr{U}$ and $\zeta: B_{b}\to\mathscr{W}$ with values in the vector spaces $\mathscr{U}$ and $\mathscr{W}$, respectively. Let $\bv{\chi}_{ab}: B_{a}\to B_{b}$ denote a diffeomorphism from $B_{a}$ to $B_{b}$. We then define the functions $\varphi^{\circd{b}}$ and $\zeta^{\circd{a}}$ as
%+-
\begin{equation*}
\label{eq: circd notation}
\varphi^{\circd{b}} \coloneqq \varphi \circ \bv{\chi}_{ab}^{-1}
\quad\text{and}\quad
\zeta^{\circd{a}} \coloneqq \zeta \circ \bv{\chi}_{ab}.
\end{equation*}
%+-
The symbols $\circinline{e}$, $\circinline{s}$, and $\circinline{f}$ are used to represent $B(t)$, $B_{\s}$, and $B_{\f}$, respectively.
\end{remark}

\begin{remark}[Concerning gradients]
The gradient and divergence operators are defined with respect to an underlying metric, which is itself associated with a particular domain. Since functions may be defined on different domains, the interpretation and practical evaluation of these operators necessarily depend on the domain over which the corresponding function is defined.    
\end{remark}

\begin{remark}[Regarding the arguments of maps]
In accordance with a common convention in Continuum Mechanics (see, e.g., Ref.~\cite{GurtinFried_2010_The-Mechanics_0}), the arguments of mappings, such as `$(\bv{X}_{\s},t)$' or `$(\bv{X}_{\f})$,' will be omitted whenever the context provides sufficient information to avoid ambiguity, thereby enabling a more concise notation.
\end{remark}

For regions away from the interface, the deformation gradients associated with the solid and fluid constituents are defined by
%+-
\begin{equation}
\label{eq: Fs and Ff defs}
    \ts{F}_{\s}(\bv{X}_{\s},t) \coloneqq \Grad{\bv{\chi}_{\s}(\bv{X}_{\s},t)}
\quad\text{and}\quad    
    \ts{F}_{\f}(\bv{X}_{\f},t) \coloneqq \Grad{\bv{\chi}_{\f}(\bv{X}_{\f},t)},
\end{equation}
%+-
with corresponding determinants denoted by $J_{\s}$ and $J_{\f}$, assumed to be strictly positive. The velocity of the solid and fluid phases are, respectively, defined as
%+-
\begin{equation}
\label{eq: vs and vf defs}
    \bv{v}_{\s}(\bv{X}_{\s},t) \coloneqq \partial_{t}\bv{\chi}_{\s}(\bv{X}_{\s},t)
    \quad\text{and}\quad
    \bv{v}_{\f}(\bv{X}_{\f},t) \coloneqq \partial_{t}\bv{\chi}_{\f}(\bv{X}_{\f},t).
\end{equation}
%+-
For each fixed time $t \in \mathbb{R}_{0}^{+}$, we define
%+-
\begin{equation}
\label{eq: Deformations defs}
\prescript{}{t}{\bv{\chi}}_{\s}(\bv{X}_{\s}) \coloneqq 
\bv{\chi}_{\s}(\bv{X}_{\s},t)
\quad\text{and}\quad
\prescript{}{t}{\bv{\chi}}_{\f}(\bv{X}_{\f}) \coloneqq
\bv{\chi}_{\f}(\bv{X}_{\f},t).
\end{equation}
Away from the interface, the mappings $\prescript{}{t}{\bv{\chi}}_{\s}(\bv{X}_{\s}): B_{\s}\to B(t)$ and $\prescript{}{t}{\bv{\chi}}_{\f}(\bv{X}_{\f}):B_{\f}\to B(t)$ are assumed to be diffeomorphism on $B_{\s}$ and $B_{\f}$, respectively. 
Assuming that the solid and fluid phases can be regarded as submanifolds embedded within a common Euclidean manifold, the displacements associated with the solid and fluid constituents, denoted by $\bv{u}_{\s}$ and $\bv{u}_{\f}$, respectively, are defined as
%+-
\begin{align}
\label{eq: us and uf defs}
    \bv{u}_{\s}(\bv{X}_{\s},t) \coloneqq \bv{\chi}_{\s}(\bv{X}_{\s},t) - \bv{X}_{\s}
    \quad\text{and}\quad
    \bv{u}_{\f}(\bv{X}_{\f},t) \coloneqq \bv{\chi}_{\f}(\bv{X}_{\f},t) - \bv{X}_{\f}.
\end{align}
%+-
The material accelerations corresponding to the solid and fluid phases, represented by $\bv{a}_{\s}$ and $\bv{a}_{\f}$, respectively, are defined as
%+-
\begin{align}
\label{eq: as and af defs}
\bv{a}_{\s}(\bv{X}_{\s},t) \coloneqq \partial_{t}\bv{v}_{\s}(\bv{X}_{\s},t)
\quad\text{and}\quad
\bv{a}_{\f}(\bv{X}_{\f},t) \coloneqq \partial_{t}\bv{v}_{\f}(\bv{X}_{\f},t).
\end{align}
%+-

The external boundary of the domain in the current configuration is denoted by $\partial B_{\text{ext}}(t)$ and corresponds, for each $t \in \mathbb{R}^{+}_{0}$, to the image of the external boundary of the solid domain in the reference configuration, $\partial B_{\s_{\text{ext}}}$, under the mapping
\begin{equation*}
\prescript{}{t}{\bv{\chi}}_{\s}(\bv{X}_{\s}) \coloneqq 
\bv{\chi}_{\s}(\bv{X}_{\s},t)
\quad \text{for}\ t \in \mathbb{R}^{+}_{0}.
\end{equation*}
The unit normal vectors associated with the external boundary of the domain in the current configuration and reference configuration of the solid are denoted by $\bv{n}$ and $\bv{n}_{\s}$, respectively.

\begin{remark}[Definition of fluid velocity field from an implementation point of view]
In the finite element implementation, the fluid velocity field is defined on the computational domain, i.e.,
%+-
\begin{equation*}
\bv{v}_{\f}=\bv{v}_{\f} \left( \bv{X}_{\s},t \right) .
\end{equation*}
%+-
Therefore, the fluid's material acceleration is described by (cf.\ \cite{Fernandez2009The-Derivation-0})
%+-
\begin{align*}
\bv{a}_{\f} &=\bv{a}_{\f} \left( \bv{X}_{\s},t \right) =\partial\bv{v}_{\f} + \ts{L}_{\f} \ts{F}_{\s}^{-1} \left( \bv{v}_{\f} - \bv{v}_{\s} \right)
\end{align*}
%+-
\end{remark} 

Regarding the interface, let $\mathcal{S}(t)$ represent a permeable interface that separates the porous domain from the purely fluid domain. For every $t$, the interface $\mathcal{S}(t)$ is taken to be the image of a smooth, non-self-intersecting fixed submanifold $\mathcal{S}_{\s}\subset B_{\s}$ under $\prescript{}{t}{\bv{\chi}}_{\s}$.
The notations $\mathcal{S}^{+}(t)$ and $\mathcal{S}^{-}(t)$, or equivalently $\mathcal{S}_{\s}^{+}$ and $\mathcal{S}_{\s}^{-}$, indicate that the interface is approached from the purely fluid domain and the porous domain, respectively.

The symbol $\bv{m}_{\s}$ is used to denote the unit normal vector that orients $\mathcal{S}_{\s}$, with the convention that it points from the porous domain toward the pure fluid domain. For each $t \in \mathbb{R}^{+}_{0}$, the associated unit normal vector orienting $\mathcal{S}(t)$ is denoted by $\bv{m}$ and is defined as
%+-
\begin{equation}
\label{eq: normal to St}
\bv{m} \coloneqq \biggl(\frac{J_{\s}\inversetranspose{\ts{F}_{\s}}\bv{m}_{\s}}{\|J_{\s}\inversetranspose{{\ts{F}_{\s}}} \bv{m}_{\s}\|}\biggr)^{\circd{e}},
\end{equation}
%+-

\section{Balance of Mass and Momentum}
\label{sec: balance of mass and momentum}

\subsection{Porous Medium}

Here, a discussion on the mass and momentum balance laws governing the physics of the porous medium is presented. The material in this Section has been developed in detail in Refs.~\cite{Bowen1976Theory-of-Mixtu0,Bowen1980Incompressible-0,Costanzo2026PoroelasticFlow}.

\subsubsection{Balance of Mass}
The fluid and solid phases are assumed to be incompressible in their pure states, with constant mass densities denoted by $\rho_{\f}^{*}$ and $\rho_{\s}^{*}$, respectively.

Let
%+-
\begin{equation}
\label{eq: volume fractions}
\phi_{\s}^{-}: B^{-}(t)\to[0,1]
\quad\text{and}\quad
\phi_{\f}^{-}: B^{-}(t)\to[0,1]
\end{equation}
%+-
represent the volume fractions associated with the solid and fluid phases, respectively. These quantities satisfy the \emph{saturation condition}
%+-
\begin{equation}
\label{eq: saturation condition}
\phi_{\s}^{-}+\phi_{\f}^{-} = 1.
\end{equation}
%+-
Accordingly, the mass densities of the solid and fluid constituents within the mixture are given, respectively, by
%+-
\begin{equation}
\label{eq: mass density distributions}
\rho_{\s}^{-} = \rho_{\s}^{*} \phi_{\s}^{-}
\quad\text{and}\quad
\rho_{\f}^{-} = \rho_{\f}^{*} \phi_{\f}^{-}.
\end{equation}
%+-
In the absence of chemical reactions within $B^{-}(t)$, conservation of mass implies that
%+-
\begin{equation}
\label{eq: balance of mass}
\partial_{t}\rho_{a}^{-} + \Div{\Bigl( \rho_{a}^{-} \left( \bv{v}_{a}^{-} \right)^{\circd{e}}\Bigr)} = 0
\quad\Rightarrow\quad
\partial_{t}\phi_{a}^{-} + \Div{\Bigl(\phi_{a}^{-} \left(\bv{v}_{a}^{-} \right)^{\circd{e}}\Bigr)} = 0,
\end{equation}
%+-
for each constituent $a = \s,\f$.
The second relation in Eqs.~\eqref{eq: balance of mass}, together with the assumed regularity of the motions away from the interface, implies that referential representations of the volume fraction distributions exist for both the solid and fluid phases. These referential volume fractions are denoted by $\phi_{R_{\s}}^{-}$ and $\phi_{R_{\f}}^{-}$, respectively:
%+-
\begin{equation}
\label{eq: phiRs and phiRf defs}
\phi_{R_{\s}}^{-} = J_{\s}^{-} \left( \phi_{\s}^{-} \right)^{\circd{s}}
\quad\text{and}\quad
\phi_{R_{\f}}^{-} = J_{\f}^{-} \left( \phi_{\f}^{-} \right)^{\circd{f}}.
\end{equation}
%+-
By employing Eq.~\eqref{eq: saturation condition}, the motion of the mixture is subject to the following single constraint:
%+-
\begin{equation}
\label{eq: div vvavg constraint}
\Div{\Bigl(\phi_{\s}^{-} \left(\bv{v}_{\s}^{-} \right)^{\circd{e}}+\phi_{\f}^{-} \left(\bv{v}_{\f}^{-} \right)^{\circd{e}}\Bigr)} = 0.
\end{equation}
%+-
The following fields are introduced over $B^{-}(t)$:
%+-
\begin{equation}
\label{eq: vvavg and vflt defs}
\bv{v}_{\va}^{-} \coloneqq \phi_{\s}^{-} \left(\bv{v}_{\s}^{-} \right)^{\circd{e}}+\phi_{\f}^{-} \left(\bv{v}_{\f}^{-} \right)^{\circd{e}}
\quad\text{and}\quad
\bv{v}_{\flt}^{-} \coloneqq \phi_{\f}^{-}\Bigl( \left(\bv{v}_{\f}^{-} \right)^{\circd{e}} - \left(\bv{v}_{\s}^{-} \right)^{\circd{e}}\Bigr).
\end{equation}
%+-
The field $\bv{v}_{\va}^{-}$ represents the volume-averaged velocity of the mixture, while $\bv{v}_{\flt}^{-}$ denotes the filtration velocity. By employing Eqs.~\eqref{eq: vvavg and vflt defs}, Eq.~\eqref{eq: div vvavg constraint} may equivalently be expressed as
%+-
\begin{equation}
\label{eq: div vvavg constraint expanded}
\Div{\bv{v}_{\va}^{-}} = 0
\quad\text{or}\quad
\Div{\left(\bv{v}_{\s}^{-}\right)^{\circd{e}}} + \Div{\bv{v}_{\flt}^{-}} = 0.
\end{equation}
%+-

\subsubsection{Balance of Momentum}

The mixture is taken to be isothermal. Following the formulations presented in Refs.~\cite{Bowen1976Theory-of-Mixtu0} and~\cite{Bowen1980Incompressible-0}, the balance of momentum for phase $a = \s,\f$ within the mixture reads, throughout $B^{-}(t)$ and away from the interface,
%+-
\begin{equation}
\label{eq: momentum}
\rho_{a}^{-} \left( \bv{a}_{a}^{-} \right)^{\circd{e}} - \Div{\ts{T}_{a}^{-}} -\bv{b}_{a}^{-} - \bv{p}_{a}^{-} = \bv{0},
\end{equation}
%+-
where $\ts{T}_{a}^{-}$ denotes the Cauchy stress associated with species $a$, $\bv{b}_{a}^{-}$ represents an external force density per unit volume of $B^{-}(t)$, and $\bv{p}_{a}^{-}$ denotes the force density per unit volume of $B^{-}(t)$ exerted on phase $a$ through its interaction with the remainder of the mixture. These interaction forces are required to satisfy the constraint
%+-
\begin{equation}
\label{eq: interaction force constraint}
\sum_{a} \bv{p}_{a}^{-} = \bv{0}.
\end{equation}
%+-
Rather than considering the equation of motion for the solid component, the equation of motion for the mixture as a whole is employed, which is obtained by summing the equations of motion of the fluid and solid components.

\subsection{Pure Fluid}

Assuming the fluid to be incompressible, the mass and momentum balance equations in the current configuration of the fluid domain $B^{+}(t)$ are given by
%+-
\begin{align}
\label{eq: fluid momentum balance}
\rho_{\f}^{+} \left( \bv{a}_{\f}^{+} \right)^{\circd{e}} - \Div{\ts{T}_{\f}^{+}} -\bv{b}_{\f}^{+}  &= \bv{0},\\
\label{eq: fluid mass balance}
\Div{\bv{v}_{\f}^{+}} &= 0,
\end{align}
%+-
where the symbol $\circinline{e}$ is used to denote the Eulerian representation of the field.
Here, $\rho_{\f}^{+}$ is the density of the pure fluid, $\ts{T}_{\f}^{+}$ denotes the Cauchy stress, and $\bv{b}_{\f}^{+}$ represents the body force per unit volume of $B^{+}(t)$.
The governing equation describing the mesh motion over the computational domain $B_{\s}^{+}$ is
%+-
\begin{equation}
\label{eq: fluid mesh motion}
-\Div{\ts{F}_{\s}^{+}\ts{S}_{\s}^{+}} = \bv{0},
\end{equation}
%+-
where $\ts{S}_{\s}^{+}$ denotes the second Piola--Kirchhoff stress tensor.

\begin{remark}[Governing equation of mesh motion]
The equation governing the mesh motion is chosen for convenience, subject to the geometric constraints the problem imposes. Although various strategies exist for implementing mesh motion, as discussed in \cite{Wick2011Fluid-Structure-0}, we adopt an implementation that is formally equivalent to a nonlinear quasi-static elasticity formulation.
\end{remark}

\subsection{Interface}
\label{sec: interface conditions}

Following Refs.~\cite{Bowen1976Theory-of-Mixtu0,Bowen1980Incompressible-0,Costanzo2026PoroelasticFlow}, and noting that the balance of mass for the solid component is trivially satisfied, the discontinuity of the fluid velocity across $\mathcal{S}(t)$ implies that the jump condition associated with the balance of mass is given by
%+-
\begin{equation}
\label{eq: jump of fluid mass}
\llbracket \rho_{\f} (\bv{v}_{\f}^{\circd{e}} - \bv{v}_{\s}^{\circd{e}}) \rrbracket \cdot \bv{m} = 0. 
\end{equation}
%+-
Assuming that the fluid outside the porous domain is identical to the fluid within it, we obtain
%+-
\begin{equation}
\label{eq: mass balance jump condition}
\llbracket \bv{v}_{\flt} \rrbracket \cdot \bv{m} = 0.
\end{equation}
%+-

\begin{remark}[Filtration velocity definition for the pure fluid]
Based on the second of Eqs. \eqref{eq: vvavg and vflt defs}, the definition of the filtration velocity in the pure fluid domain simplifies to
\begin{equation*}
\bv{v}_{\flt}^{+} = \left( \bv{v}_{\f}^{+} \right)^{\circd{e}} - \left( \bv{v}_{\s}^{+} \right)^{\circd{e}}
\end{equation*}
since the solid velocity has been assumed to be continuous across the interface.
\end{remark}
For the mixture momentum balance jump condition across $\mathcal{S}(t)$, we assume that
%+-
\begin{equation}
\label{eq: jump of momentum}
\llbracket
\rho_{a}(\bv{v}_{\f}^{\circd{e}} - \bv{v}_{\s}^{\circd{e}}) \otimes (\bv{v}_{\f}^{\circd{e}} - \bv{v}_{\s}^{\circd{e}}) - \ts{T}_{\s} - \ts{T}_{\f}
\rrbracket \, \bv{m}
= \bv{0},
\end{equation}
%+-
where it is recalled that $\bv{v}_{\s}$ is globally continuous.

\begin{remark}[Jump condition for the energy balance]
Because isothermal conditions are assumed, neither the energy balance nor its corresponding jump condition needs to be considered.
\end{remark}

Following the variational formulation developed in \cite{Costanzo2026PoroelasticFlow} and under the assumption of a fully dissipative interface, the jump condition associated with the fluid balance of momentum is expressed as
%+-
\begin{equation}
\label{eq: final abstract form of the jump condition of the fluid momentum}
\Bigl(
k_{\f}\bv{m}
- d \bv{v}_{\f}^{\circd{e}}
+
\phi_{\f} \ip \bv{m} + 
\ts{T}_{\f}\bv{m}
-
\tfrac{1}{2} \phi_{\f} \mu_{\mathcal{S}}
\llbracket \bv{v}_{\flt}\rrbracket
\Bigr)^{\pm}
= \bv{0} 
\quad
\text{on $\mathcal{S}(t)$}.
\end{equation}
%+-
Here, $k_{\f}=\tfrac{1}{2} \rho_{\f} \| \bv{v}_{\f}^{\circd{e}} \|^{2}$ denotes the fluid kinetic energy per unit volume in the current configuration, $\mu_\mathcal{S}$ represents the interface viscosity, having units of dynamic viscosity per unit length, $\ip$ denotes the Lagrange multiplier introduced to enforce the jump condition associated with the balance of mass, and $d$ is defined as
%+-
\begin{equation}
d =
{\rho^{*}_{\f}\phi_\f} (\bv{v}_{\f}^{\circd{e}} - \bv{v}_{\s}^{\circd{e}}) \cdot \bv{m}.
\end{equation}
%+-
It should be noted that, in the pure fluid domain, $\phi_{\f}=1$ and $\rho_{\f}^{*}=\rho_{\f}$.
For a complete derivation, refer to \cite{Costanzo2026PoroelasticFlow}.

\section{Constitutive Response Functions}
\label{sec: constitutive response functions}

\subsection{Porous Medium}

The pure solid phase is assumed to be incompressible and hyperelastic. Its strain energy is taken to be a non-negative scalar function of the form
%+-
\begin{equation}
\label{eq: pure solid strain energy}
\hat{\Psi}_{\s}^{-} = \hat{\Psi}_{\s}^{-}(\overline{\ts{C}}_{\s}^{-},\bv{X}_{\s}),
\end{equation}
%+-
where
%+-
\begin{equation}
\label{eq: right CG strain tensor}
\ts{C}_{\s}^{-} \coloneqq \transpose{(\ts{F}_{\s}^{-})} \ts{F}_{\s}^{-}
\quad\text{and}\quad
\overline{\ts{C}}_{\s}^{-} = (J_{\s}^{-})^{-2/3} \ts{C}_{\s}^{-},
\end{equation}
%+-
such that $\overline{\ts{C}}_{\s}^{-}$ represents the isochoric component of $\ts{C}_{\s}^{-}$. The strain energy of the solid phase within the mixture is assumed to be the strain energy of the pure solid multiplied by its volume fraction:
%+-
\begin{equation}
\label{eq: solid strain energy}
\Psi_{\s}^{-} \coloneqq \phi_{R_{\s}}^{-} \hat{\Psi}_{\s}^{-}.
\end{equation}
%+-
For convenience, the following notation is introduced for the solid strain energy per unit volume of the current configuration:
%+-
\begin{equation}
\label{eq: eulerian strain energy}
\psi_{\s}^{-} \coloneqq \Bigl((J_{\s}^{-})^{-1}\Psi_{\s}^{-}\Bigr)^{\circd{e}}.
\end{equation}
%+-
The elastic strain energy contributions to the first Piola--Kirchhoff and Cauchy stress tensors are given, respectively, by
%+-
\begin{equation}
\label{eq: 1PK and Cauchy stress}
\ts{P}_{\s}^{e^{-}} = 2 \phi_{R_{\s}}^{-} \ts{F}_{\s}^{-} \frac{\partial \hat{\Psi}_{\s}^{-}}{\partial \ts{C}_{\s}^{-}}
\quad\text{and}\quad
\ts{T}_{\s}^{e^{-}} = \biggl(2 \phi_{R_{\s}}^{-} (J_{\s}^{-})^{-1} \ts{F}_{\s}^{-} \frac{\partial \hat{\Psi}_{\s}^{-}}{\partial \ts{C}_{\s}^{-}} \transpose{(\ts{F}_{\s}^{-})}\biggr)^{\circd{e}}.
\end{equation}
%+-
Besides the hyperelastic contribution, a dissipative stress contribution arising from the interaction between the solid and fluid phases is also considered. This contribution is analogous in form to that presented in Eq.~(2.1.3) of Ref.~\cite{Bowen1976Theory-of-Mixtu0}. Accordingly, for $a = \s,\f$, we define
%+-
\begin{equation}
\label{eq: definition of Da}
\ts{D}_{a}^{-} \coloneqq \tfrac{1}{2} \bigr(\ts{L}_{a}^{-} + \transpose{(\ts{L}_{a}^{-})}\bigl),\quad\text{with}\quad
\ts{L}_{a}^{-} \coloneqq \Grad \left({\bv{v}_{a}^{-}}\right)^{\circd{e}}.
\end{equation}
%+-
The viscous contribution to the solid stress within the mixture is then expressed as
%+-
\begin{equation}
\label{eq: solid viscous Cauchy stress}
\ts{T}_{\s}^{v^{-}} =  2 \mu_{B} (\ts{D}_{\s}^{-} - \ts{D}_{\f}^{-}).
\end{equation}
%+-
A contribution of this form is commonly encountered in poroelasticity, although it is more often introduced in the context of the fluid response, where it is known as a Brinkman term \cite{Brinkman1949A-Calculation-o}. The dynamic viscosity $\mu_{B}$ is, in general, a function of $\phi_{\s}^{-}$ whose support coincides with that of $\phi_{\s}^{-}$.

For the pure fluid phase, incompressibility has been assumed. In addition, the fluid is taken to be linearly viscous, with viscous stress given by $2 \hat{\mu}_{\f} \ts{D}_{\f}$, where $\hat{\mu}_{\f} \in \mathbb{R}^{+}_{0}$ denotes a constant dynamic shear viscosity. The stress response of the fluid within the mixture is assumed to take the form
%+-
\begin{equation}
\label{eq: fluid stress response}
\ts{T}_{\f}^{{v,I}^{-}} = 2 \mu_{\f}^{-} \ts{D}_{\f}^{-},
\end{equation}
%+-
where the superscript `$I$' denotes `intrinsic,' and $\mu_{\f}^{-}$ is a dynamic viscosity as a function of $\phi_{\f}^{-}$. In addition to $\ts{T}_{\f}^{{v,I}^{-}}$, the corresponding Brinkman contribution introduced previously is also included (see again Eq.~(2.1.3) in Ref.~\cite{Bowen1976Theory-of-Mixtu0}). Consequently, the overall viscous stress response of the fluid is assumed to be
%+-
\begin{equation}
\label{eq: viscous Cauchy stress of fluid in mixture}
\ts{T}_{\f}^{v^{-}} = \ts{T}_{\f}^{{v,I}^{-}} + 2 \mu_{B} (\ts{D}_{\f}^{-} - \ts{D}_{\s}^{-}).
\end{equation}
%+-
The dynamic viscosity $\mu_{\f}^{-}$ is assumed to satisfy the physically intuitive limits $\mu_{\f}^{-}\to 0$ as $\phi_{\f}^{-}\to 0$ and $\mu_{\f}^{-}\to \hat{\mu}_{\f}$ as $\phi_{\f}^{-}\to 1$.

A pressure field (see, e.g., Ref.~\cite{Bowen1980Incompressible-0}), denoted by $p^{-}$ and referred to as the \emph{pore pressure}, is introduced over $B^{-}(t)$ to enforce Eq.~\eqref{eq: div vvavg constraint}. The complete expressions for $\ts{T}_{\s}^{-}$ and $\ts{T}_{\f}^{-}$ are therefore given by
%+-
\begin{equation}
\label{eq: Ts and Tf overall}
\ts{T}_{\s}^{-} = -\phi_{\s}^{-} p^{-} \, \ts{I} + \ts{T}_{\s}^{e^{-}} +  \ts{T}_{\s}^{v^{-}}
\quad\text{and}\quad
\ts{T}_{\f}^{-} = -\phi_{\f}^{-} p^{-} \, \ts{I} + \ts{T}_{\f}^{v^{-}},
\end{equation}
%+-
where $\ts{I}$ denotes the identity tensor.

The constitutive description of the mixture must also specify the quantities $\bv{p}_{a}^{-}$ for $a = \s,\f$. In both instances, these forces are assumed to arise from the viscous resistance opposing the relative motion of the fluid with respect to the solid, namely from the kinematic quantity $(\bv{v}_{\f}^{-} - \bv{v}_{\s}^{-})$.
With $\bv{p}_{\s}^{-} = -\bv{p}_{\f}^{-}$ by Eq.~\eqref{eq: interaction force constraint}, we follow Ref.~\cite{Bowen1980Incompressible-0} and adopt the expression
%+-
\begin{equation}
\label{eq: pf and ff full form}
\bv{p}_{\f}^{-} = p^{-} \Grad{\phi_{\f}^{-}} + \bv{f}_{\f},
\quad\text{with}\quad
\bv{f}_{\f} = -\frac{\mu_{D} (\phi_{\f}^{-})^{2}}{\kappa_{\s}} \Bigl( (\bv{v}_{\f}^{-})^{\circd{e}} - (\bv{v}_{\s}^{-})^{\circd{e}} \Bigr),
\end{equation}
%+-
where $\kappa_{\s}$ denotes the permeability of the solid, while $\mu_{D}$ is a dynamic viscosity that, similarly to $\mu_{B}$, has support identical to that of $\phi_{\s}^{-}$.

\subsection{Pure Fluid}

The fluid is assumed to be incompressible, and its viscous stress is expressed as
%+-
\begin{equation}
\label{eq: viscous Cauchy stress of pure fluid}
\ts{T}_{\f}^{v^{+}} = 2 \mu_{f}^{+} \ts{D}_{\f}^{+},
\end{equation}
%+-
where $\mu_{\f}^{+}$ denotes the dynamic viscosity and is equal to $\hat{\mu}_{\f}$.
By introducing the pressure field $p^{+}$ over $B^{+}(t)$, the total Cauchy stress $\ts{T}_{\f}^{+}$ of the pure fluid is given by
%+-
\begin{equation}
\ts{T}_{\f}^{+} = -p^{+} \, \ts{I} + \ts{T}_{\f}^{v^{+}}.
\end{equation}
%+-

\begin{remark}[Cauchy stress for the solid phase in the pure fluid domain]
Because the solid phase is absent from the pure fluid domain, we set
\begin{equation*}
\ts{T}_{\s}^{e^{+}}=\ts{0} \quad\text{and}\quad \ts{T}_{\s}^{v^{+}}=\ts{0}.
\end{equation*}
\end{remark}

A pseudo-strain energy (i.e., not based on physical considerations) governing the mesh motion is introduced and assumed to be a smooth convex function of the form
%+-
\begin{equation}
\label{eq: mesh motion strain energy}
\hat{\Psi}_{\s}^{+} = \hat{\Psi}_{\s}^{+}(\ts{C}_{\s}^{+},\bv{X}_{\s}).
\end{equation}
%+-
where
%+-
\begin{equation}
\label{eq: right CG strain tensor - mesh motion}
\ts{C}_{\s}^{+} \coloneqq \transpose{(\ts{F}_{\s}^{+})} \ts{F}_{\s}^{+}.
\end{equation}
%+-
For future use, we denote by $\overline{\ts{C}}_{\s}^{+}$ the isochoric component of $\ts{C}_{\s}^{+}$:
%+-
\begin{equation}
\label{eq: isochoric right CG strain tensor - mesh motion}
\overline{\ts{C}}_{\s}^{+} = (J_{\s}^{+})^{-2/3} \ts{C}_{\s}^{+}.
\end{equation}
%+-
From Eq.~\eqref{eq: mesh motion strain energy}, the mesh motion's contribution to the (pseudo-)second Piola--Kirchhoff stress is
%+-
\begin{equation}
\label{eq: 2PK stress}
\ts{S}_{\s}^{+} = 2 \frac{\partial \hat{\Psi}_{\s}^{+}}{\partial \ts{C}_{\s}^{+}}
\end{equation}
%+-

\section{Weak Formulation}
\label{sec: weak formulation}

Considering the first of Eqs. \eqref{eq: div vvavg constraint expanded} and Eq. \eqref{eq: momentum}, which represents the balance of momentum for the mixture and the fluid in the porous medium, together with Eqs. \eqref{eq: fluid mass balance}, \eqref{eq: fluid momentum balance}, and \eqref{eq: fluid mesh motion} for the pure fluid, the following variables are selected as the primary unknowns for the porous medium and the pure fluid, respectively,
%+-
\begin{equation*}
\label{eq: principal unknowns}
\bv{u}_{\s}^{-},\ \left( \bv{v}_{\f}^{-} \right)^{\circd{s}},\ \left( p^{-} \right)^{\circd{s}},
\quad\text{and}\quad
\bv{u}_{\s}^{+},\ \left( \bv{v}_{\f}^{+} \right)^{\circd{s}},\ \left( p^{+} \right)^{\circd{s}}.
\end{equation*}
%=-
Moreover, the Lagrange multiplier $\ip^{\circd{s}}$ is treated as an unknown defined on the interface.
The \ac{ALE} scheme (cf. \cite{Fernandez2009The-Derivation-0}) is employed, in which the computational domain for the porous medium is taken as the initial configuration of the solid phase, coinciding with the reference configuration of the solid component. For the pure fluid, the computational domain is chosen as the initial configuration of the fluid domain, which coincides with the reference configuration of the fluid.

We first define the functional spaces appropriate for the primary unknowns, which provide the natural framework for analyzing the well-posedness of the problem. For the sake of simplicity, we focus on regularity properties in the interior of the domain and along the interface, leaving aside issues related to boundary and interface conditions.
Establishing the well-posedness of the problem analytically is beyond the scope of this paper. Nevertheless, we demonstrate empirically that the proposed functional spaces yield a stable solution to the problem.

Considering the global continuity requirement of the displacement field, the following functional space is introduced for this field:
%+-
\begin{align}
\label{equ: us functional space}
\mathcal{V}^{\bv{u}_\s} & \coloneqq \left\{\bv{u}_\s \in \left[L^2\left(B_\s\right)\right]^d \mid
\Grad{\bv{u}_\s} \in \left[L^{\infty}\left(B_\s\right)\right]^{d \times d},\text{~$u_{\s_{i}} = u_{\g_{i}}$ on $\Gamma_{\s_{i}}^{D_{\bv{u}_{\s}}}$} \right\},
\end{align}
%+-
where $B_{\s}=B_{\s}^{-} \cup B_{\s}^{+}$, and $\Gamma_{\s_{i}}^{D_{\bv{u}_{\s}}}$ denotes the portion of the boundary in the reference configuration of the solid component on which the Dirichlet boundary condition $u_{\g_{i}}$ is prescribed for the $i$th component of the displacement field $u_{\s_{i}}$.

We assume that the fluid velocity is discontinuous across the interface; therefore, the following function spaces are introduced for the fluid phase of the mixture and for the pure fluid:
%+-
\begin{align}
\label{equ: vf- functional space}
\mathcal{V}^{\bv{v}_\f^{-}} & \coloneqq \left\{\bv{v}_\f \in \left[L^2\left(B^{-}(t)\right)\right]^d \mid
\Grad
{\bv{v}_\f} \in \left[L^2\left(B^{-}(t)\right)\right]^{d \times d},
\text{~$\bv{v}_{\f} = \bv{v}_{\g}^{-}$ on $\Gamma_{\f}^{D_{\bv{v}_{\f}}^{-}}$}
\right\},\\
\label{equ: vf+ functional space}
\mathcal{V}^{\bv{v}_\f^{+}} & \coloneqq \left\{\bv{v}_\f \in \left[L^2\left(B^{+}(t)\right)\right]^d \mid
\Grad
{\bv{v}_\f} \in \left[L^2\left(B^{+}(t)\right)\right]^{d \times d},
\text{~$\bv{v}_{\f} = \bv{v}_{\g}^{+}$ on $\Gamma_{\f}^{D_{\bv{v}_{\f}}^{+}}$}
\right\},
\end{align}
%+-
where $\Gamma_{\f}^{D_{\bv{v}_{\f}}^{-}}$ and $\Gamma_{\f}^{D_{\bv{v}_{\f}}^{+}}$ denote the portions of the boundary of the computational domains corresponding to the porous medium and the pure fluid, respectively, on which the Dirichlet boundary conditions $\bv{v}_{\g}^{-}$ and $\bv{v}_{\g}^{+}$ are prescribed.

For the pore pressure $p^{-}$ and the fluid pressure field $p^{+}$, we introduce the following functional spaces:
%+-
\begin{align}
\label{equ: p- functional space}
\mathcal{V}^{p^{-}} & \coloneqq \left\{p \in \left[L^2\left(B^{-}(t)\right)\right] \mid
\Grad
{p}
\in \left[L^2\left(B^{-}(t)\right)\right]^{d}
\right\},\\
\label{equ: p+ functional space}
\mathcal{V}^{p^{+}} & \coloneqq \left\{p \in \left[L^2\left(B^{+}(t)\right)\right] \mid
\Grad
{p}
\in \left[L^2\left(B^{+}(t)\right)\right]^{d}
\right\}.
\end{align}
%+-
We also define the following spaces for the test functions:
%+-
\begin{equation}
\begin{aligned}
\label{equ: test functional space}
\mathcal{V}_{0}^{\bv{u}_\s} & \coloneqq \left\{\bv{u}_\s \in \left[L^2\left(B_\s\right)\right]^d \mid
\Grad{\bv{u}_\s} \in \left[L^{\infty}\left(B_\s\right)\right]^{d \times d},\text{~$u_{\s_{i}} = 0$ on $\Gamma_{\s_{i}}^{D_{\bv{u}_{\s}}}$} \right\},\\
\mathcal{V}_{0}^{\bv{v}_\f^{-}} & \coloneqq \left\{\bv{v}_\f \in \left[L^2\left(B^{-}(t)\right)\right]^d \mid
\Grad
{\bv{v}_\f} \in \left[L^2\left(B^{-}(t)\right)\right]^{d \times d},
\text{~$\bv{v}_{\f} = \bv{0}$ on $\Gamma_{\f}^{D_{\bv{v}_{\f}}^{-}}$}
\right\},\\
\mathcal{V}_{0}^{\bv{v}_\f^{+}} & \coloneqq \left\{\bv{v}_\f \in \left[L^2\left(B^{+}(t)\right)\right]^d \mid
\Grad
{\bv{v}_\f} \in \left[L^2\left(B^{+}(t)\right)\right]^{d \times d},
\text{~$\bv{v}_{\f} = \bv{0}$ on $\Gamma_{\f}^{D_{\bv{v}_{\f}}^{+}}$}
\right\},\\
\mathcal{V}_{0}^{p^{-}} & \coloneqq \left\{p \in \left[L^2\left(B^{-}(t)\right)\right] \mid
\Grad
{p}
\in \left[L^2\left(B^{-}(t)\right)\right]^{d}
\right\},\\
\mathcal{V}_{0}^{p^{+}} & \coloneqq \left\{p \in \left[L^2\left(B^{+}(t)\right)\right] \mid
\Grad
{p}
\in \left[L^2\left(B^{+}(t)\right)\right]^{d}
\right\}.
\end{aligned}
\end{equation}
%+-

Considering the absence of contributions from the elasticity and Brinkman terms in the pure fluid domain, the jump condition for the mixture balance of momentum, Eq. \eqref{eq: jump of momentum}, can be rewritten as
%+-
\begin{equation}
\begin{aligned} 
    \llbracket
    \rho_{\f}(\bv{v}_{\f}^{\circd{e}} - \bv{v}_{\s}^{\circd{e}}) \otimes (\bv{v}_{\f}^{\circd{e}} - \bv{v}_{\s}^{\circd{e}})
    \rrbracket \bv{m}
    &= \llbracket \ts{T}_{\s} + \ts{T}_{\f} \rrbracket \bv{m} \\
    &= \Bigl[\bigl(\ts{T}^{+}_{\s} + \ts{T}^{+}_{\f}\bigr) - \bigl(\ts{T}^{-}_{\s} + \ts{T}^{-}_{\f}\bigr)\Bigr] \bv{m} \\
    &=
    \begin{multlined}[t]
        \Bigl[\bigl(
        \underbrace{\ts{T}_{\s}^{e^+}}_{= \ts{0}} + \underbrace{\ts{T}_{\s}^{v^+}}_{= \ts{0}} +
        \ts{T}_{\f}^{v^{+}} - p^{+}\ts{I}\bigr)\Bigr.\\
        - \Bigl.\bigl( \ts{T}_{\s}^{e^{-}} + \ts{T}_{\s}^{v^{-}} +
        \ts{T}_{\f}^{v^{-}} - p^{-}\ts{I}\bigr)\Bigr] \bv{m}.
    \end{multlined}
\end{aligned}
\end{equation}
%+-
Therefore, we obtain
\begin{equation}
\label{eq: mixture jump reformulation}
\begin{aligned}
    -\bigl[\ts{T}_{\s}^{e^{-}} + \ts{T}_{\s}^{{v}^{-}} + \ts{T}_{\f}^{v^{-}} \bigr] \bv{m}
    &= \llbracket
    \rho_{\f}(\bv{v}_{\f}^{\circd{e}} - \bv{v}_{\s}^{\circd{e}}) \otimes (\bv{v}_{\f}^{\circd{e}} - \bv{v}_{\s}^{\circd{e}}) +p\ts{I}
    \rrbracket \bv{m}
    - \ts{T}_{\f}^{v^{+}} \bv{m},
\end{aligned}
\end{equation}
which represents the traction applied by the pure fluid to the porous medium across the interface.
Since the solid displacement field is assumed to remain continuous across the interface, no additional interface condition is required to govern the mesh motion within the pure fluid domain.

The continuity of the normal component of the filtration velocity, which follows from the mass balance, is imposed by means of a Lagrange multiplier belonging to the following functional space, for both the trial and test functions:
%+-
\begin{equation}
\mathcal{V}^{\ip} \coloneqq \left\{\ip \in \left[L^2\left(\mathcal{S}(t)\right)\right] \right\}.
\end{equation}
%+-
By taking into account the continuity of the solid velocity field across the interface alongside Eq. \eqref{eq: mixture jump reformulation}, the following weak formulation is proposed for the problem:
\begin{equation}
\label{eq: weak form with new jump}
\begin{aligned}[b]
% LINE
    0 &= \int_{B_{\s}^{-}}
    J_{\s} \bigl(
    \rho_{\s} \bv{a}_{\s}^{\circd{e}} + \rho_{\f} \bv{a}_{\f}^{\circd{e}} - \bv{b}_{\s} - \bv{b}_{\f} + \Grad{p}
    \bigr)^{\circd{s}} 
    \cdot \tilde{\bv{u}}_{\s}/\tau
    + \int_{B_{\s}^{-}}
    J_{\s}
    (\ts{T}_{\s}^{e} + \ts{T}_{\s}^{v} + \ts{T}_{\f}^{v})^{\circd{s}} \inversetranspose{\ts{F}}_{\s} \colondot \Grad{\tilde{\bv{u}}_{\s}}/\tau
% LINE
\\
    &\quad
    - \int_{\partial B^{-}_{\s_{\text{ext}}}} \left( \ts{T}_{\s}^{e} + \ts{T}_{\s}^{v} + \ts{T}_{\f}^{v} \right)^{\circd{s}} J_{\s} \inversetranspose{\ts{F}}_{\s}  \bv{n}_{\s}
    \cdot \tilde{\bv{u}}_{\s}/\tau
% LINE
\\
    &\quad+
    \int_{B_{\s}^{-}}
    J_{\s} \bigl[
    (\rho_{\f} \bv{a}_{\f}^{\circd{e}} - \bv{b}_{\f} + \phi_{\f}\Grad{p} - \bv{f}_{\f})^{\circd{s}} \cdot \tilde{\bv{v}}_{\f}^{\circd{s}}
    + (\ts{T}_{\f}^{v})^{\circd{s}} \inversetranspose{\ts{F}}_{\s} \colondot \Grad{\tilde{\bv{v}}_{\f}^{\circd{s}}}\bigr]
    - \int_{\partial B^{-}_{\s_{\text{ext}}}} J_{\s} (\ts{T}_{\f}^{v})^{\circd{s}} \inversetranspose{\ts{F}}_{\s} \bv{n}_{\s}
    \cdot \tilde{\bv{v}}_{\f}^{\circd{s}}
% LINE
\\
    &\quad
    -
    \int_{B_{\s}^{-}}
    J_{\s} (\bv{v}_{\s} + \bv{v}_{\flt}^{\circd{s}}) \cdot \inversetranspose{\ts{F}}_{\s}\Grad{\tilde{p}^{\,\circd{s}}}
    +
    \int_{\partial B_{\s_{\text{ext}}}^{-}} \tilde{p}^{\circd{s}} (\bv{v}_{\s} + \bv{v}_{\flt}^{\circd{s}}) \cdot J_{\s} \inversetranspose{\ts{F}}_{\s} \bv{n}_{\s}
% LINE
\\
    &\quad
    + \int_{B^{+}(t)}\bigl[
    (\rho_{\f} \bv{a}_{\f}^{\circd{e}} - \bv{b}_{\f} + \Grad{p}) \cdot \tilde{\bv{v}}_{\f}
    + \ts{T}_{\f}^{v} \colondot \Grad{\tilde{\bv{v}}_{\f}} \bigr]
    - \int_{\partial B_{\text{ext}}^{+}(t)} \ts{T}_{\f}^{v} \bv{n} \cdot \tilde{\bv{v}}_{\f}
% LINE
\\
    &\quad
    - \int_{B^{+}(t)} \Grad{\tilde{p}_{\f}} \cdot \bv{v}_{\f}
    + \int_{\partial B^{+}_{\text{ext}}(t)} \tilde{p} \bv{v}_{\f} \cdot \bv{n}
% LINE
\\
    &\quad
    + \int_{B^{+}_{\s}} \ts{S}_{\s} \colondot \tilde{\ts{E}}_{\s}/\tau
    - \int_{\partial B^{+}_{\s_\text{ext}}}
    \ts{F}_{\s}\ts{S}_{\s}^{e} \bv{n}_{\s} \cdot \tilde{\bv{u}}_{\s}/\tau
    - \int_{\mathcal{S}_{\s}^{+}} \ts{F}_{\s} \ts{S}_{\s}^{e} \bv{n}_{\s} \cdot \tilde{\bv{u}}_{\s}/\tau
% LINE
\\
    &\quad+
    \int_{\mathcal{S}_{\s}}
    \llbracket
    \rho_{\f}(\bv{v}_{\f}^{\circd{e}} - \bv{v}_{\s}^{\circd{e}}) \otimes (\bv{v}_{\f}^{\circd{e}} - \bv{v}_{\s}^{\circd{e}}) + p \ts{I}
    \rrbracket^{\circd{s}} J_{\s} \inversetranspose{\ts{F}}_{\s}  \bv{m}_{\s}
    \cdot \tilde{\bv{u}}_{\s}/\tau
    - \int_{\mathcal{S}^{+}(t)} \ts{T}_{\f}^{v} \bv{m} \cdot \left( \tilde{\bv{u}}_{\s}/\tau \right)^{\circd{e}}
% LINE
\\
    &\quad-
    \int_{\mathcal{S}_{\s}}
    \biggl\llbracket
    \Bigl(
    k_{\f} \ts{I}
    -
    \rho_{\f} \bv{v}_{\f}^{\circd{e}} \otimes (\bv{v}_{\f}^{\circd{e}} - \bv{v}_{\s}^{\circd{e}})
    \Bigr)^{\circd{s}} J_{\s} \inversetranspose{\ts{F}}_{\s} \bv{m}_{\s} \cdot \tilde{\bv{v}}_{\f}^{\circd{s}}
    \biggr\rrbracket
% LINE
\\
    &\quad-
    \int_{\mathcal{S}_{\s}}
    \Bigl\llbracket
    %\Bigl[
    \bigl(
    \phi_{\f} (\ip - p) \, \ts{I}
    -
    \tfrac{1}{2} \phi_{\f} \mu_{\mathcal{S}} \llbracket \bv{v}_{\flt} \rrbracket 
    \otimes \bv{m}
    \bigr)^{\circd{s}} J_{\s}\inversetranspose{\ts{F}}_{\s}\bv{m}_{\s}
    \cdot \tilde{\bv{v}}_{\f}^{\circd{s}}
    \Bigr\rrbracket
% LINE
\\
    &\quad
    - \int_{\mathcal{S}_{\s}}
    \llbracket
    (\bv{v}_{\s} +
    \bv{v}_{\flt}^{\circd{s}}) \, \tilde{p}^{\circd{s}}
    \rrbracket \cdot J_{\s} \inversetranspose{\ts{F}}_{\s} \bv{m}_{\s}
    +
    \int_{\mathcal{S}_{\s}} 
    \llbracket
    \bv{v}_{\flt}^{\circd{s}}
    \rrbracket
    \tilde{\ip}^{\circd{s}} \cdot
    J_{\s} \inversetranspose{\ts{F}}_{\s} \bv{m}_{\s}
\end{aligned}
\end{equation}
where the subscript $\text{ext}$ denotes the external boundary of each domain, and $\tau$ is a unit time constant introduced to ensure dimensional consistency. The divergence theorem is applied only to the continuity equation in both domains, whereas it is not applied to the pressure term in the balance equations. Furthermore, to simplify the notation as much as possible, the $+$ and $-$ superscripts have been omitted, and the integration boundaries explicitly indicate the domain or boundary over which each integral is to be evaluated.

\section{Results}
\label{sec: results}

This section begins with a numerical verification of the approach, carried out using the method of manufactured solutions. Then, a representative problem is solved in order to examine how the interface properties influence the coupled behavior of the problem.

All numerical implementations described here rely on \comsol, employed as the \ac{FE} programming environment. Coupling between the equations is handled through a monolithic solver, and time is advanced using a variable time step, variable order \ac{BDF} scheme.
Standard \ac{FEM} along with \ac{ALE} formulations is used to implement Eq.~\eqref{eq: weak form with new jump}. A conformal representation was adopted for the interface $\mathcal{S}_{\s}$, meaning it was constructed from inter-element boundaries. The interpolation functions selected for each field are the following:
    \begin{itemize}
        \item The globally continuous displacement field is represented with \nth{2} order Lagrange polynomials.
        \item The \nth{2} order Lagrange polynomials are used for the fluid velocity fields of both the porous medium and the free fluid, as these fields are discontinuous across the interface.
        \item The field $p$ is discontinuous across the interface and is interpolated with \nth{1} order Lagrange polynomials.
        \item The field $\ip$, defined solely on the interface, is likewise interpolated using \nth{1} order Lagrange polynomials.
    \end{itemize}

\subsection{Verification}

\begin{figure}[h]
    \centering
    \includegraphics[width=0.3\linewidth]{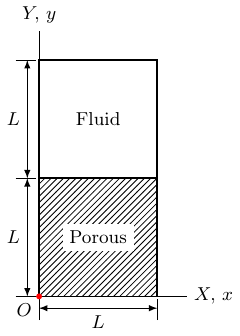}
    \caption{Problem Setup}
    \label{fig: mms problem setup}
\end{figure}

The formulation is verified using the method of manufactured solutions \cite{Salari2000Code-Verification-0}. Figure~\ref{fig: mms problem setup} shows the two-dimensional geometry considered for this purpose: a pure fluid domain sits atop a poroelastic domain. 
%+-
%+-
As shown in Fig.~\ref{fig: mms problem setup}, points in the reference configuration are located by the position $\bv{X}$, given by coordinates $X$ and $Y$. In the deformed configuration, the position $\bv{x}$ of these points is described by the coordinates $x = X + u_{s_X}(X, Y, t)$ and $y = Y + u_{s_Y}(X, Y, t)$, with $u_{s_X}$ and $u_{s_Y}$ representing the displacement components in the horizontal and vertical directions, respectively.
Under this motion, the pure fluid and porous domains map to $B^{+}(t)$ and $B^{-}(t)$, respectively, while the interface's reference configuration $\mathcal{S}_{\s}$ is given by the segment with $X\in[0,L]$ and $Y = L$.

For the strain energy density of the pure solid phase in the porous domain, a neo-Hookean hyperelastic model is adopted, $\hat{\Psi}_{\s}^{-}=\tfrac{1}{2} \mu_{R_e} \left(\bar{I}_1^{-} - 3 \right)$, with $\bar{I}_1^{-}$ denoting the first principal invariant of the isochoric part of the right Cauchy--Green strain tensor $\bar{\ts{C}}_{\s}^{-}$, and $\mu_{R_e}$ the referential solid's elastic shear modulus. In addition, $\mu_{\f}^{-}=\bar{\mu}_\f \left( \phi_{\f}^{-} \right)^2$ and $\mu_B=\bar{\mu}_B \phi_\f^2$ are assumed to hold.
The mesh motion in the pure fluid domain uses a compressible Yeoh hyperelastic model for its energy density function, $\hat{\Psi}_{\s}^{+}=c_{1} \left(\bar{I}_1^{+} - 3 \right) + c_{2} \left(\bar{I}_1^{+} - 3 \right)^{2} + \tfrac{1}{2} K (J_{\s}^{+} - 1)^{2} $, where $c_{1}=\np[Pa]{1}$, $c_{2}=\np[Pa]{2.5}$, and $K=\np[Pa]{1}$. Here too, $\bar{I}_1^{+}$ is the first principal invariant of the isochoric part of the right Cauchy--Green strain tensor $\bar{\ts{C}}_{\s}^{+}$, and $\mu_{\f}^{+}=\bar{\mu}_\f$ holds.
Material properties for both domains are listed in Table~\ref{tab:mms-properties}. The values chosen for most parameters draw on biological applications \cite{Kedarasetti2022Arterial-vasodi}.
%+-
\begin{table}[hbt]
\centering
\caption{Parameter values employed in the convergence rate study}
\label{tab:mms-properties}
\begin{tabular}{cccc}
\toprule
Quantity & Pure Fluid Domain & Porous Domain & Interface \\
\midrule
$\rho_s^*$          &                      & \np[kg/m^3]{1e3}     &                       \\
$\mu_{R_e}$         &                      & \np[Pa]{5e3}         &                       \\
$\kappa_{\s}$       &                      & \np[m^2]{1e-9}       &                       \\
$\phi_{R_{\s}}$     &                      & $0.4$                &                       \\
$\mu_D$             &                      & \np[Pa\cdot s]{1e-3} &                       \\
$\bar{\mu}_B$       &                      & \np[Pa\cdot s]{1e-3} &                       \\
$\bar{\mu}_\f$      & \np[Pa\cdot s]{1e-3} & \np[Pa\cdot s]{1e-3} &                       \\
$\rho_f^*$          & \np[kg/m^3]{1e3}     & \np[kg/m^3]{1e3}     &                       \\
$\mu_\mathcal{S}$   &                      &                      & \np[Pa\cdot s/m]{1e2} \\
\bottomrule
\end{tabular}
\end{table}
%+-

\begin{figure}[hbt]
    \centering
    \begin{subfigure}{0.15\textwidth}
        \centering
        \includegraphics[width=\linewidth]{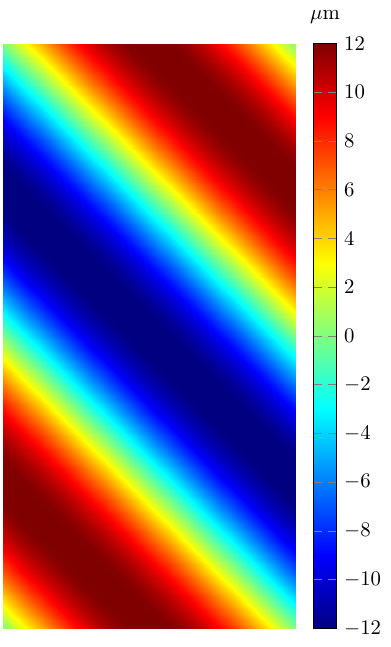}
        \caption{$\bv{u}_{\s_X}$.}
    \end{subfigure}
    \begin{subfigure}{0.15\textwidth}
        \centering
        \includegraphics[width=\linewidth]{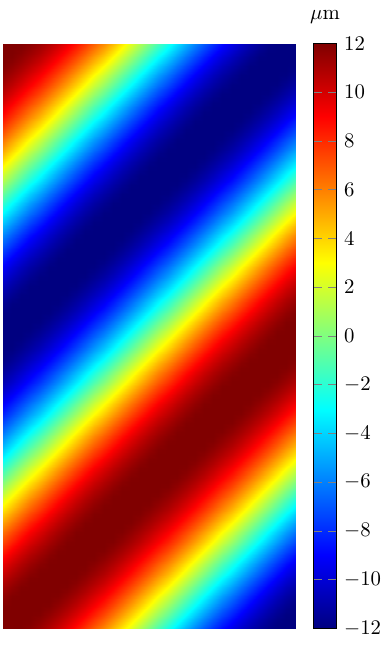}
        \caption{$\bv{u}_{\s_Y}$.}
    \end{subfigure}
    \\
    \begin{subfigure}{0.15\textwidth}
        \centering
        \includegraphics[width=\linewidth]{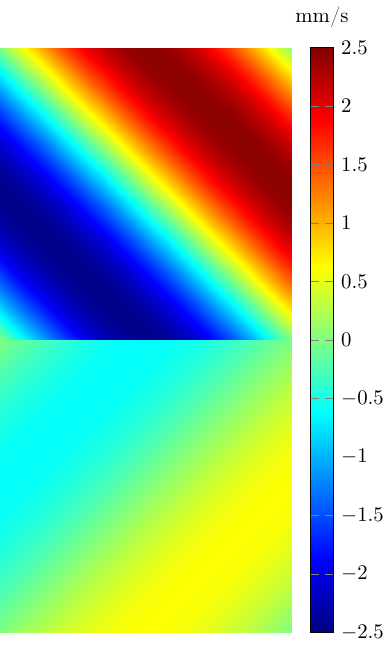}
        \caption{$\bv{v}_{\f_x}^{\circd{s}}$.}
    \end{subfigure}
    \begin{subfigure}{0.15\textwidth}
        \centering
        \includegraphics[width=\linewidth]{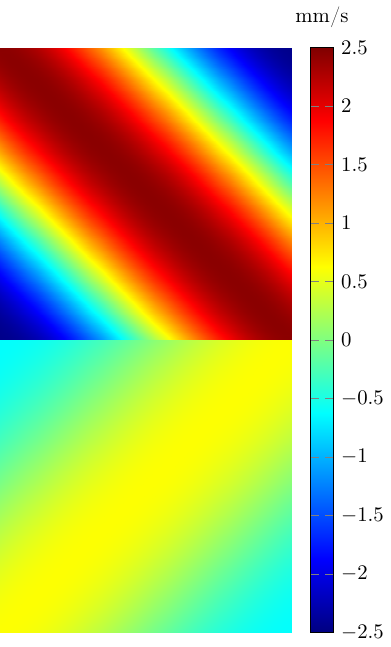}
        \caption{$\bv{v}_{\f_y}^{\circd{s}}$.}
    \end{subfigure}
    \begin{subfigure}{0.15\textwidth}
        \centering
        \includegraphics[width=\linewidth]{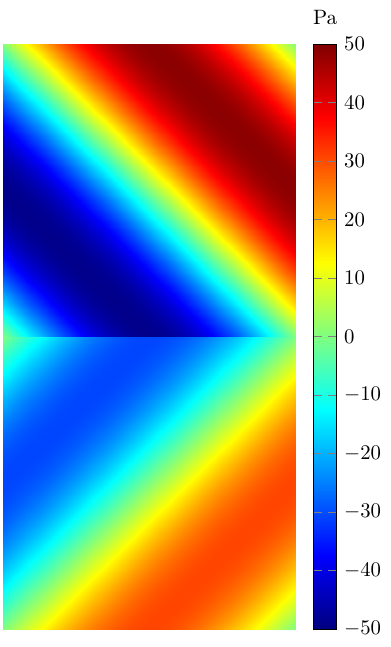}
        \caption{$p^{\circd{s}}$.}
    \end{subfigure}
    \caption{Panels (a) and (b) present the manufactured solutions corresponding to the displacement components of the solid phase. Panels (c) and (d) depict the velocity components of the fluid phase, while panel (e) illustrates the pressure field. All quantities are shown at time $t=\np[s]{0.1}$. For clarity, every plot is displayed with respect to the reference configuration of the solid phase.}
    \label{fig:mms-fields}
\end{figure}

The manufactured solution is composed of the following fields:
%+-
\begin{align}
    %% Pore pressure in +
    \label{eq: mms pA}
    p^{+} &= p_o^{+} \bigl(1-\cos(\pi t/t_o)\bigr) \sin\bigl[\pi(x+y)/L\bigr]
        & &\text{in $B^{+}(t)$},
    \\
    %% Pore pressure in -
    \label{eq: mms pB}
    p^{-} &= p_o^{-} \bigl(1-\cos(\pi t/(2t_o))\bigr) \sin\bigl[\pi(x-y)/L\bigr]
        & &\text{in $B^{-}(t)$},
    \\
    %% Solid displacement X component
    \label{eq: mms usXAB}
    u_{\s_{X}} &=
        u_o \bigl(1-\cos(\pi t/(2t_o))\bigr) \sin\bigl[\pi(X+Y)/L\bigr]
            & &\text{in $B_{\s}^{+}\cup B_{\s}^{-}$},
    \\
    %% Solid displacement Y component
    \label{eq: mms usYAB}
    u_{\s_{Y}} &=
        u_o \bigl(1-\cos(\pi t/(2t_o))\bigr) \cos\bigl[\pi(X-Y)/L\bigr]
            & &\text{in $B_{\s}^{+}\cup B_{\s}^{-}$},
    \\
    %% Fluid velocity in + x component
    \label{eq: mms vfxA}
    v_{\f_{x}}^{+} &= v_{o}^{+} \bigl(1-\cos(\pi t/t_o)\bigr) \sin\bigl[\pi(x+y)/L\bigr]
        & &\text{in $B^{+}(t)\cup B^{-}(t)$,}
    \\
    %% Fluid velocity in + y component
    \label{eq: mms vfyA}
    v_{\f_{y}}^{+} &=  v_{o}^{+} \bigl(1-\cos(\pi t/t_o)\bigr) \cos\bigl[\pi(x+y)/L\bigr]
        & &\text{in $B^{+}(t)\cup B^{-}(t)$,}
    \\
    %% Fluid velocity in - x component
    \label{eq: mms vfxB}
    v_{\f_{x}}^{-} &= v_{o}^{-} \bigl(1-\cos(\pi t/(2t_o))\bigr) \sin\bigl[\pi(x-y)/L\bigr]
        & &\text{in $B^{+}(t)\cup B^{-}(t)$,}
    \\
    %% Fluid velocity in - y component
    \label{eq: mms vfyB}
    v_{\f_{y}}^{-} &= v_{o}^{-} \bigl(1-\cos(\pi t/(2t_o))\bigr) \cos\bigl[\pi(x-y)/L\bigr]
        & &\text{in $B^{+}(t)\cup B^{-}(t)$,}
    \\
    %% interface pressure
    \label{eq: mms ipSt}
    \ip &= \ip_o \bigl(1-\cos(\pi t/t_o)\bigr) \cos(\pi x/L)
        & &\text{on $\mathcal{S}(t)$}.
\end{align}
%+-
Here $u_{o}=\np[mm]{1}$, $v_{o}^{+}=\np[mm/s]{50}$, $p_{o}^{+}=\np[kPa]{1}$, $v_{o}^{-}=\np[mm/s]{50}$, $p_{o}^{-}=\np[kPa]{2.5}$, $\ip_{o}=\np[kPa]{1}$, $L=\np[mm]{10}$, and $t_{o}=\np[s]{1}$.
The chosen manufactured solution is illustrated in Fig.~\ref{fig:mms-fields} in terms of solid displacement, fluid velocity, and pressure.
Substituting these fields into the strong form of the governing equations yields the source terms needed both in the interiors of their respective domains and on the interface.
%+-
%+-

The manufactured solution, evaluated at $t = 0$, determines the initial conditions. As the boundary conditions, displacements are prescribed across the entire external boundary, with the exception of the porous domain's bottom boundary, where an elastic traction is applied instead. Moreover, the filtration velocity is imposed on the porous domain's external boundary; the lateral edges of the pure fluid domain carry prescribed velocity; and traction is applied along the pure fluid domain's top boundary.

Time integration relied on a second-order \ac{BDF} scheme with a maximum constant time step of $\np[s]{5e-4}$. Newton's method, with an automatically determined damping factor, was used to solve the nonlinear system, and the corresponding linear system was resolved with the MUMPS direct solver. Termination was governed by the smaller of the solution-based and residual-based estimated relative errors, with all solver parameters left at their \comsol default values; these correspond to a relative tolerance of $10^{-3}$ and a residual scaling factor of $10^{3}$, noting that \comsol applies its own particular definition of stopping criteria.

A mesh of identical triangles was used, refined uniformly over five successive cycles from an initial mesh diameter of $h = \np[mm]{2.5}$.

\begin{figure}[h]
    \centering
    \begin{subfigure}[b]{0.45\textwidth}
        \centering
        \includegraphics[width=\linewidth]{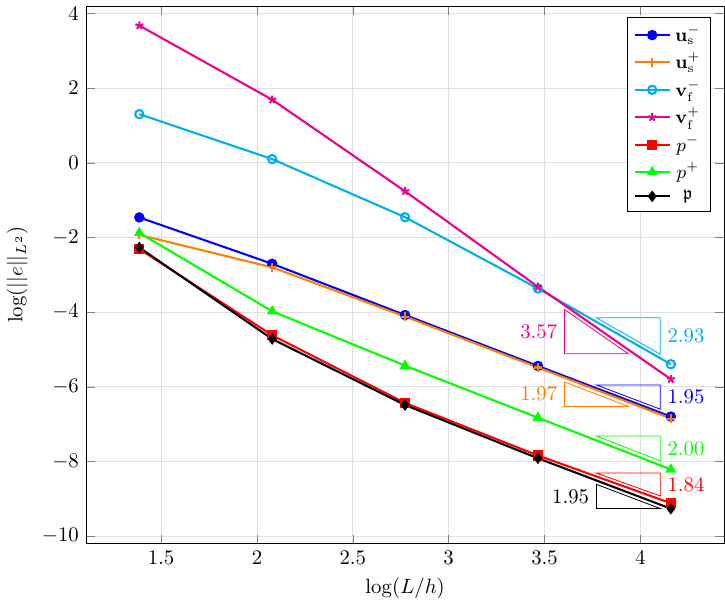}
        \caption{$L^{2}$ error}
        \label{fig: mms L2 convergence rate}
    \end{subfigure}
    \hfill
    \begin{subfigure}[b]{0.45\textwidth}
        \centering
        \includegraphics[width=\linewidth]{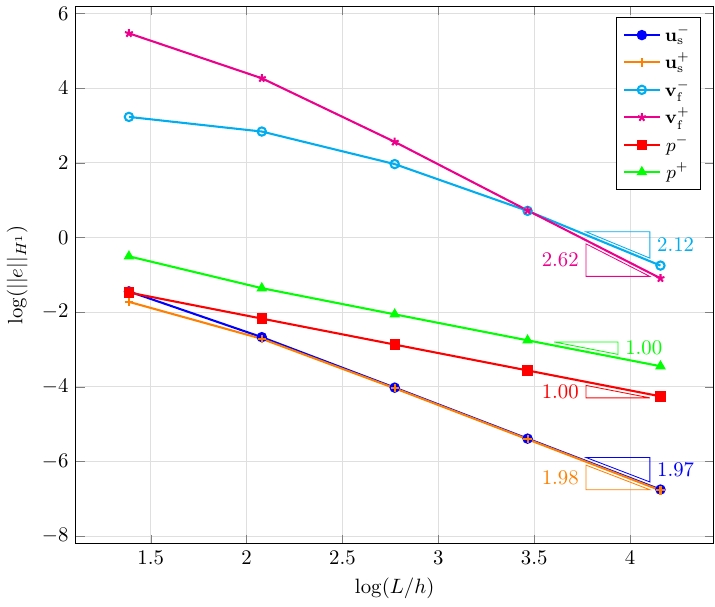}
        \caption{$H^{1}$ error}
        \label{fig: mms H1 convergence rate}
    \end{subfigure}
    \caption{Convergence behavior is reported in \subref{fig: mms L2 convergence rate} for the $L^2$-norm and in \subref{fig: mms H1 convergence rate} for the $H^1$-semi-norm of the relative error $e$ between the numerical and exact solutions, across all variables, at time $t=\np[s]{0.1}$. Norms of the relative errors are computed through integrals that respect each error's domain of definition.}
    \label{fig: mms convergence rate}
\end{figure}

The convergence rates obtained under uniform refinement are shown in Fig.~\ref{fig: mms convergence rate}, reporting the $L^2$-norm and $H^1$-semi-norm of the relative errors for each quantity appearing in the manufactured solution. Since the functions in Eqs.~\eqref{eq: mms pA}--\eqref{eq: mms ipSt} are each defined over their own particular domain, computing the corresponding relative errors requires integrals consistent with those domains.
%+-
%+-
In the absence of a formal analysis to establish whether these convergence rates are optimal, the rates observed here compare favorably with those typically reported for the Navier--Stokes problem.

\subsection{Mixed Poiseuille--Couette Flow}

We validated our numerical framework by comparing its results with the analytical solution of the mixed Poiseuille--Couette flow problem under the classical interface conditions developed by Hou et al. \cite{Hou1989BoundaryConditi}. Here, a viscous fluid moves over a porous medium, driven jointly by a constant pressure gradient and the motion of an impermeable moving wall at the top. Since the bottom wall is rigid and impermeable, the fluid velocity vanishes there. Both within the channel and inside the porous medium, the flow is unidirectional, oriented along the $x$ direction.
Hou et al. \cite{Hou1989BoundaryConditi} treated the Poiseuille and Couette flow cases separately, developing interface boundary conditions between a porous medium and an adjoining pure fluid. Their approach rested on a ``pseudo-no-slip'' kinematic condition, motivated by the requirement that interface behavior between mixtures, or between a mixture and a fluid, recover the standard single-phase continuum result. A well-posed formulation for each subdomain, they noted, demands more than satisfying the mass and momentum jump conditions alone; additional kinematic constraints must also be imposed. The constraints they settled on were continuity of the pore pressure together with full continuity of the filtration velocity.
%+-
\begin{figure}[htb]
    \centering
    \includegraphics[width=0.5\linewidth]{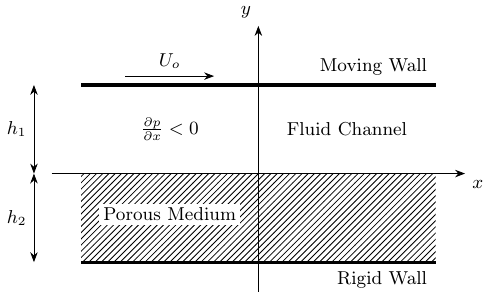}
    \caption{This benchmark problem, known as the Mixed Poiseuille--Couette flow setup, considers a viscous fluid moving across a porous domain under the combined action of a constant pressure gradient and a moving wall.}
    \label{fig: benchmark problem setup}
\end{figure}
%+-
Since the flow is unidirectional and the porous medium is assumed to be rigid, the mass balance holds trivially, and the momentum balance reduces, in dimensionless form, to the following equations for the channel and the porous medium, respectively:
\begin{align}
    \dfrac{\partial^2 V_{\f}^{+}}{\partial Y^2}+1 = 0, \quad & \text{for $0 < Y < H$},
    \shortintertext{and}
    \delta^2 \dfrac{\partial^2 V_{\f}^{-}}{\partial Y^2} - V_{\f}^{-} + \eta \, \delta^2 = 0, \quad & \text{for $-1 < Y < 0$},
\end{align}
with the second equation obtained by simplifying Eq.~\eqref{eq: momentum} for the mixture, and the first corresponding to the dimensionless Navier--Stokes equation with inertia and body forces neglected.
The dimensionless fluid velocities in the channel and the porous medium are denoted by
\begin{gather}
V_{\f}^{+}(Y)=\frac{\mu_{\f} v_{\f_{x}}^{+}}{h_{2}^{2} \left( -\frac{\partial p}{\partial x} \right)}
\shortintertext{and}
V_{\f}^{-}(Y)=\frac{\phi_{\f} \mu_{\f} v_{\f_{x}}^{-}}{h_{2}^{2} \left( -\frac{\partial p}{\partial x} \right)}, 
\end{gather}
respectively. The horizontal component of the fluid velocity is indicated by the subscript $x$.
The remaining dimensionless parameters are defined as $Y = y/h_{2}$, $H = h_{1}/h_{2}$, $\eta = \phi^{2}_{\f} \mu_{\f}/\mu_{\mathrm{a}}$, and $\delta = \sqrt{\mu_{\mathrm{a}}/(h_{2}^{2} \kappa_{\s})}$, with $\mu_{\mathrm{a}}$ denoting the apparent viscosity of the fluid within the porous medium.
Because the bottom boundary is impermeable and fixed, the fluid velocity is set to zero there, whereas at the top boundary the dimensionless fluid velocity equals unity.
The interface conditions governing the interaction between a porous medium and a free fluid, taken from \cite{Hou1989BoundaryConditi}, are stated as
\begin{gather}
 \phi_{\f} v_{\f_{x}}^{-} \Big|_{y=0^{-}} = v_{\f_{x}}^{+} \Big|_{y=0^{+}}
 \shortintertext{and}
 \mu_{\mathrm{a}} \frac{\partial v_{\f_{x}}^{-}}{\partial y} \Big|_{y=0^{-}}
 = \phi_{\f} \mu_{\f} \frac{\partial v_{\f_{x}}^{+}}{\partial y} \Big|_{y=0^{+}}.
\end{gather}
Therefore, the dimensionless form of the interface conditions is given as
\begin{gather}
 V_{\f}^{-} \Big|_{Y=0^{-}} = V_{\f}^{+} \Big|_{Y=0^{+}}
 \shortintertext{and}
  \frac{\partial V_{\f}^{-}}{\partial Y} \Big|_{Y=0^{-}}
 = \eta\, \frac{\partial V_{\f}^{+}}{\partial Y} \Big|_{Y=0^{+}}.
\end{gather}

\begin{figure}[hbt]
    \centering
    \begin{subfigure}[b]{0.48\linewidth}
        \centering
        \includegraphics[width=\linewidth]{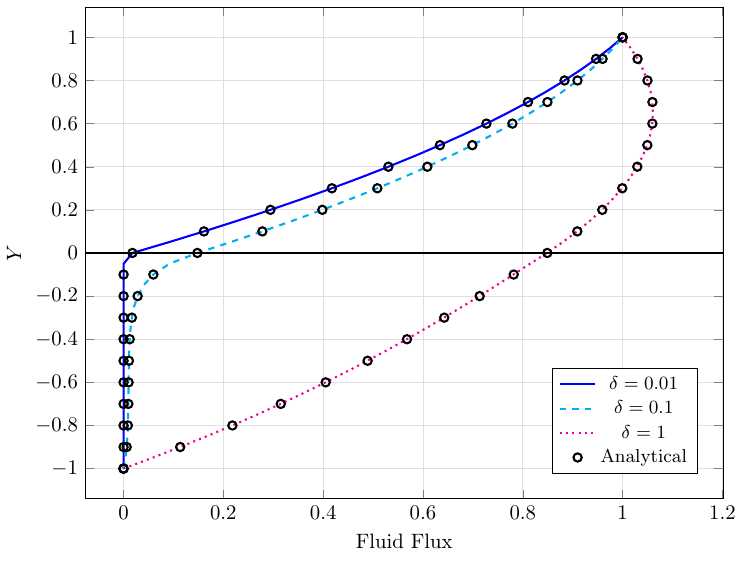}
        \caption{$H=\np{1}$}
        \label{fig: benchmark problem results h1}
    \end{subfigure}
    \hfill
    \begin{subfigure}[b]{0.48\linewidth}
        \centering
        \includegraphics[width=\linewidth]{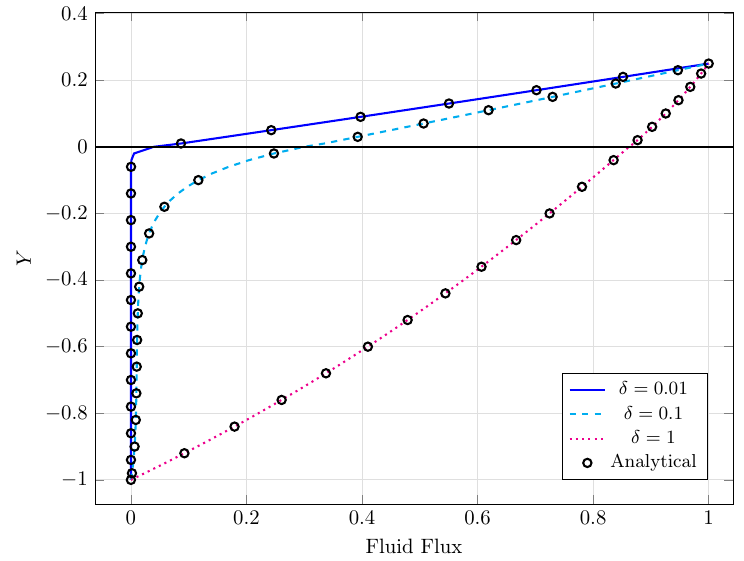}
        \caption{$H=\np{0.25}$}
        \label{fig: benchmark problem results h0.25}
    \end{subfigure}
    \caption{Fluid flux profiles are shown for the porous medium and the fluid channel at $\eta=1$, for various values of $\delta$, with panel \subref{fig: benchmark problem results h1} corresponding to $H=1$ and panel \subref{fig: benchmark problem results h0.25} to $H=0.25$. The lines depict results from the formulation developed in this paper, whereas the hollow circles represent the analytical solution obtained using the jump conditions from \cite{Hou1989BoundaryConditi}.}
    \label{fig: benchmark problem results}
\end{figure}

We reproduced the analytical solution using the stationary form of our \ac{FE} model, adopting the parameter values from Table~\ref{tab:mms-properties} with one exception: the interface viscosity $\mu_{\mathcal{S}}$ was set to $\np[Pa \cdot s/m]{1e3}$ so as to fully enforce continuity of the filtration velocity. To recover the one-dimensional character of the problem exactly, periodic conditions were imposed on the lateral boundaries.
Figure~\ref{fig: benchmark problem results} presents both the analytical solution and the dimensionless results produced by the numerical model. 
%+-
%+-
Despite the absence of any elasticity contribution in the analytical solution, the model reproduces it successfully. This is because the porous medium's deformation is negligible in this particular problem, the solid's elastic shear modulus being high enough relative to the internal forces present.

\subsection{Sample Problem}
\label{sec: results sample problem}

\begin{figure}[hbt]
\centering
\includegraphics[width=0.8\linewidth]{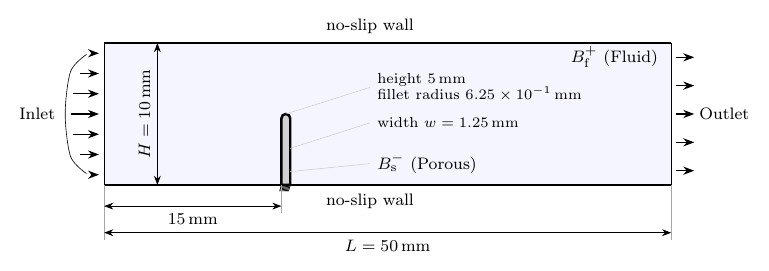}
\caption{A thin biphasic porous structure sits within a fluid channel in this sample problem, exchanging mass and momentum with the surrounding fluid across the permeable interface. The velocity of the fluid takes a parabolic form at the inlet and its pressure is set to zero at the outlet. At the bottom, the porous structure's boundary is impermeable and fully fixed in both degrees of freedom.}
\label{fig: sample problem setup}
\end{figure}

A representative test case, widely used in \ac{FSI} research, is employed here to examine the influence of the interface model on the coupled response. The configuration, depicted in Fig.~\ref{fig: sample problem setup}, is two-dimensional and couples the deformation of a biphasic porous structure with the flow of a viscous fluid through a channel.
The domain takes the form of a horizontal rectangular channel with length $L=\np[mm]{50}$ and height $H=\np[mm]{10}$. Clamped to the channel's bottom wall, $\np[mm]{15}$ from the inlet, is a slender body measuring $\np[mm]{1.25}$ in width and $\np[mm]{5}$ in length. Uniformity in the out-of-plane direction is assumed throughout.
Material properties follow those listed in Table~\ref{tab:mms-properties}, with the exception of $\phi_{R_\s}=0.7$. Starting from a quasi-static state, the simulation imposes fixity and impermeability along the bottom edge of the porous domain.
No-slip conditions govern the channel's top and bottom boundaries, while zero pressure is prescribed at the outlet. At the inlet, the fluid velocity is imposed according to the profile
\begin{equation*}    
v_{\textrm{in}}=v_{o} \frac{y}{H} \left( 1- \frac{y}{H} \right) \left( 1 - \cos \frac{\pi t}{t_{o}} \right).
\end{equation*}
In this expression, $v_{o}=\np[mm/s]{200}$, $t_{o}=\np[s]{4}$, and the total simulation duration is $\np[s]{4}$. 
%+-
%+-
The solver settings match those employed for verification via the method of manufactured solutions, apart from the maximum time step, set here to $\np[s]{1e-2}$.
The domain is discretized using identical triangular elements of size $h=\np[mm]{1.56e-1}$, supplemented at the interface by $\np{8}$ boundary layers with a stretching factor of $\np{1.2}$.

\begin{figure}[hbt]
    \centering
    \includegraphics[width=0.7\linewidth]{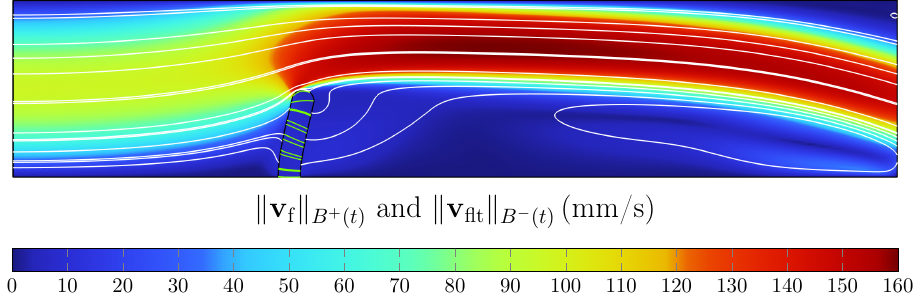}
    \caption{The fluid passes into the channel through the left boundary, substantially deforming the thin biphasic structure as it moves through and exits at the right boundary.
    The fluid velocity and its streamlines (white) in the pure fluid domain, as well as the filtration velocity and its streamlines (green) in the porous domain, are shown.
    Because the interface viscosity $\mu_{\mathcal{S}}$ is not comparatively large, the tangential component of the filtration velocity is not tightly enforced to be continuous, which is why a jump in the tangential filtration velocity is visible at the interface.}
    \label{fig: sample problem fluid velocity}
\end{figure}

\begin{figure}[hbt]
    \centering
    \begin{subfigure}[b]{0.49\linewidth}
        \centering
        \includegraphics[height=0.75\linewidth]{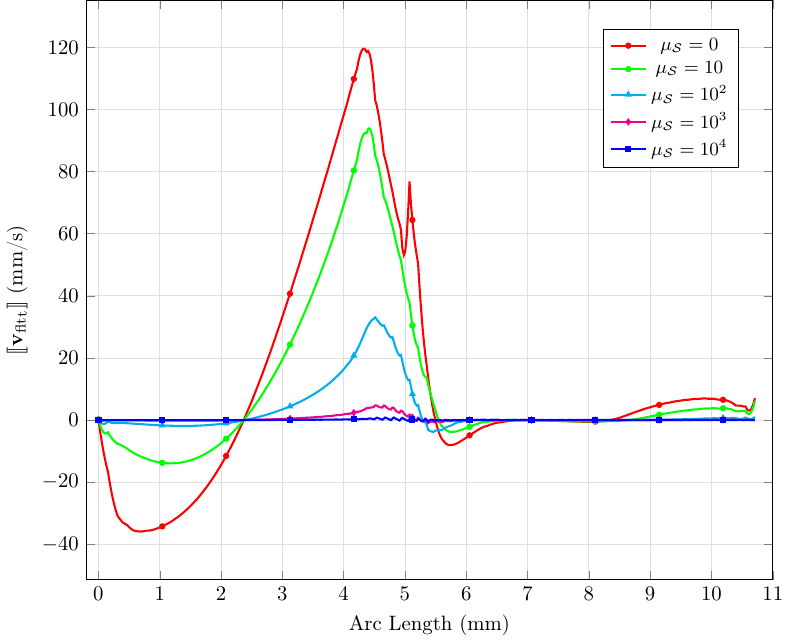}
        \caption{}
        \label{fig: sample problem filt vel tan jump}
    \end{subfigure}
    \hfill    
    \begin{subfigure}[b]{0.49\linewidth}
        \centering
        \includegraphics[height=0.75\linewidth]{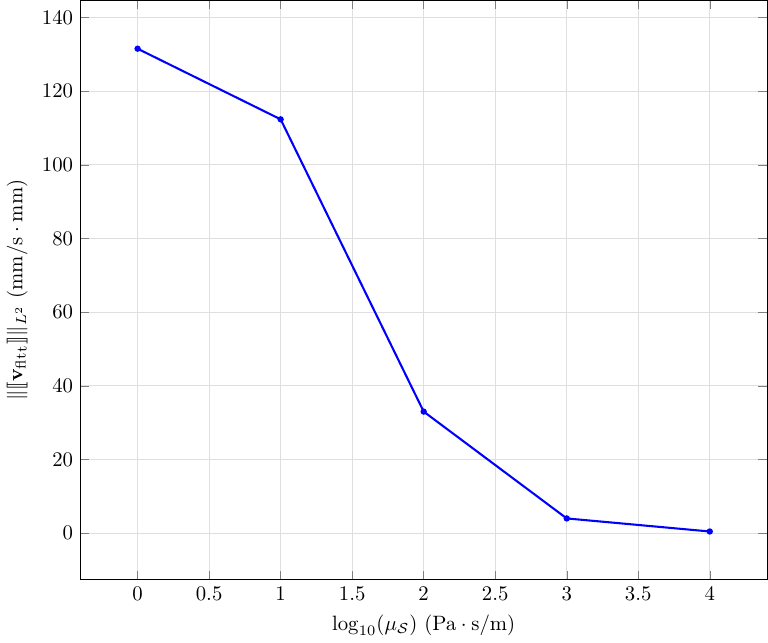}
        \caption{}
        \label{fig: sample problem filt vel tan jump norm}
    \end{subfigure}
    \caption{Panel~\subref{fig: sample problem filt vel tan jump} displays the jump in the filtration velocity tangential component, while panel~\subref{fig: sample problem filt vel tan jump norm} presents the corresponding $L^{2}$ norm of the jump for $\bv{v}_{\flt_{\mathrm{t}}}$. As $\mu_{\mathcal{S}}$ increases, continuity of the tangential filtration velocity component is progressively restored.}
    \label{fig: sample problem filt vel}
\end{figure}

\begin{figure}
    \centering
    \includegraphics[width=0.5\linewidth]{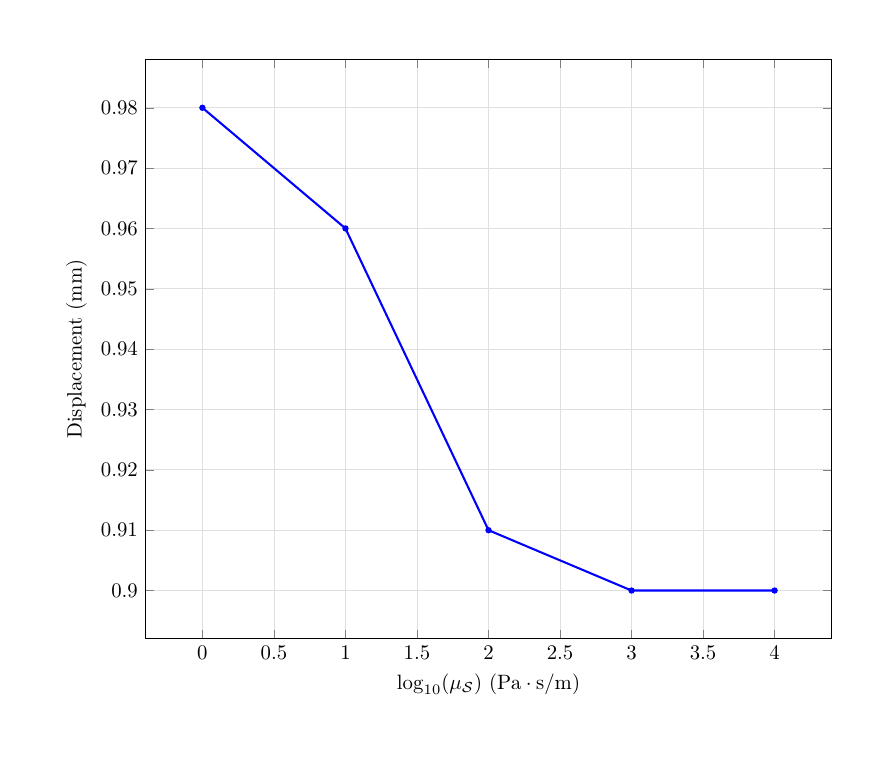}
    \caption{The tip point's displacement in the porous structure is shown for various values of the interface viscosity; as this viscosity grows, the structure grows relatively more rigid.}
    \label{fig: sample problem tip disp}
\end{figure}

The magnitudes of the fluid velocity field in the pure fluid domain and the filtration velocity field in the porous domain at $t=\np[s]{4}$, when the porous body undergoes substantial deformation, are shown in Fig.~\ref{fig: sample problem fluid velocity}.
%+-
%+-
A discontinuity in fluid velocity exists at the interface; nevertheless, enforcing continuity of the tangential component of the filtration velocity is possible depending on the value of the interface viscosity $\mu_{\mathcal{S}}$. Figure~\ref{fig: sample problem filt vel tan jump} shows the jump of the tangential component of the filtration velocity at the interface across different values of the interface viscosity. As this viscosity grows larger, continuity in the filtration velocity is achieved and the norm of the jump in the tangential component of the filtration velocity goes to zero, as shown in Fig.~\ref{fig: sample problem filt vel tan jump norm}.
%+-
%+-
Indeed, the tangential component of the filtration velocity becomes continuous, given that continuity of its normal component is already guaranteed by the Lagrange multiplier $\ip$, which arises from enforcing mass balance continuity at the interface.
Figure~\ref{fig: sample problem tip disp} illustrates how the interface condition enforcing continuity of the tangential filtration velocity component affects the results, showing the tip displacement of the thin porous structure as a function of the interface viscosity $\mu_{\mathcal{S}}$.
%+-
%+-

\section{Discussion and Conclusion}
\label{sec: discussion and conclusion}

This work introduced an \ac{ALE} framework capable of modeling \ac{FPSI} between a free fluid and a porous structure described through mixture theory. Building on interface conditions we had previously derived for the shared boundary between two contiguous poroelastic domains, we reformulated these conditions for the boundary between a free fluid and a poroelastic medium. A single, continuous displacement field underlies the approach, simultaneously capturing the motion of the mixture's solid constituent and the \ac{ALE} map governing the deforming pure fluid domain. For the interface itself, a fully dissipative character was assumed, though alternative interface models remain possible.

The numerical examples presented offer encouraging support for the method: convergence of our formulation was demonstrated empirically via the method of manufactured solutions, a process that also pointed to a practical choice of stable finite element spaces. Additionally, by solving the classic problem of flow through a rigid porous medium, we recovered several well-established results, achieved by tuning the constitutive response prescribed at the interface. The formulation includes a dissipative term tied to the norm of the jump in filtration velocity, which can be understood as penalizing that jump. Increasing the parameter $\mu_{\mathcal{S}}$ causes the solution to approach one in which the filtration velocity is continuous.

This study pursues two objectives. The first is to demonstrate that the model we developed earlier extends naturally to \ac{FSI} modeling and can reproduce outcomes equivalent to the traditional jump conditions, including full continuity of the filtration velocity. The second is to illustrate how modeling assumptions can carry hidden, intrinsic consequences. In fact, assuming a fully dissipative interface together with a large interface viscosity allowed us to recreate the results yielded by the traditional jump conditions, which suggests that those conditions implicitly encode a hidden physical assumption. Consequently, care should be taken with the assumptions adopted, so as not to impose constraints that have no basis in the underlying physics of the problem.

Motivated by brain physiology and the behavior of membranes within the central nervous system, we intend to pursue applications in this direction in future work; in particular, the present framework offers a means of probing physical assumptions made in the brain biomechanics model described in \cite{Jannesari2026ModificationOfA}. Such applications will call for more elaborate models governing the pore pressure jump at the interfaces involved. Earlier studies enforced continuity of the pressure field as an interface condition (cf. \cite{Shim2021FiniteElementIm}); by contrast, the present framework permits the pressure field to be discontinuous, which makes it suitable for settings such as osmosis, where a genuine discontinuity in the interfacial pressure field is physically required.

The present work does not address the micromechanical origins of the interface behavior or how such properties might be translated into a constitutive law. Determining this constitutive behavior---particularly when the interface model targets a specific application such as brain biomechanics---could be pursued through homogenization techniques or informed directly by experimental characterization.
Another shortcoming of the present formulation concerns the limiting case in which porosity vanishes near the interface. In this regime, the model should recover the classical \ac{FSI} problem, yet the current interface conditions fail to remain well-posed, so no results can be produced in this limit. Addressing this would require pairing the modification proposed by our previous study \cite{Jannesari2026ModificationOfA} for the poroelastic domain itself with a corresponding adjustment to the interface conditions, so that the coupled formulation remains valid as this extreme limit is approached.

\section*{Acknowledgments}
The authors gratefully acknowledge partial support from the Pennsylvania Department of Health using Tobacco CURE Funds (the Department specifically disclaims responsibility for any analyses, interpretations, or conclusions).

\bibliography{Literature}

@incollection{Bowen1976Theory-of-Mixtu0,
	address = {New York},
	author = {Bowen, R. M.},
	booktitle = {Continuum Physics},
	editor = {Eringen, A. C.},
	pages = {1--127},
	publisher = {Academic Press},
	title = {Theory of Mixtures},
	volume = {III---Mixtures and EM Field Theories},
	year = {1976}}

@article{Bowen1980Incompressible-0,
	author = {Bowen, R. M.},
	journal = {International Journal of Engineering Science},
	note = {{DOI: 10.1016/0020-7225(80)90114-7}},
	number = {9},
	pages = {1129--1148},
	title = {Incompressible Porous Media Models by Use of the Theory of Mixtures},
	volume = {18},
	year = {1980}}

@article{dellIsola2009Boundary-Condit-0,
	author = {dell'Isola, F. and Madeo, A. and Seppecher, P.},
	journal = {International Journal of Solids and Structures},
	note = {{DOI: 10.1016/j.ijsolstr.2009.04.008}},
	number = {17},
	pages = {3150--3164},
	title = {Boundary Conditions at Fluid-Permeable Interfaces in Porous Media: A Variational Approach},
	volume = {46},
	year = {2009}}

@book{GurtinFried_2010_The-Mechanics_0,
	address = {New York},
	author = {Gurtin, M. E. and Fried, E. and Anand, L.},
	publisher = {Cambridge University Press},
	title = {The Mechanics and Thermodynamics of Continua},
	year = {2010}}

@Article{Costanzo2026PoroelasticFlow,
  author    = {Costanzo, Francesco and Jannesari, Mohammad and Ghitti, Beatrice},
  journal   = {Mathematical Models and Methods in Applied Sciences},
  title     = {Poroelastic flow across a permeable interface: a {H}amilton’s principle approach and its finite element implementation},
  year      = {2026},
  issn      = {1793-6314},
  month     = mar,
  doi       = {10.1142/s0218202526500259},
  publisher = {World Scientific Pub Co Pte Ltd},
}

@article{Brinkman1949A-Calculation-o,
	author = {Brinkman, H. C.},
	journal = {Applied Scientific Research},
	note = {{DOI: 10.1007/BF02120313}},
	pages = {27--34},
	title = {A Calculation of the Viscous Force Exerted by a Flowing Fluid on a Dense Swarm of Particles},
	volume = {1},
	year = {1949}}

@incollection{Fernandez2009The-Derivation-0,
	author = {Fern{\'a}ndez, M. A. and Formaggia, L. and Gerbeau, J-.F. and Quarteroni, A.},
	booktitle = {Cardiovascular Mathematics: Modeling and Simulation of the Circulatory System},
	chapter = {3},
	editor = {Formaggia, L. and Quarteroni, A. and Veneziani, A.},
	note = {{DOI: 10.1007/978-88-470-1152-6\_3}},
	pages = {77--121},
	publisher = {Springer},
	series = {Modeling, Simulation and Applications (MS\&A)},
	title = {The Derivation of the Equations for Fluids and Structure},
	volume = {1},
	year = {2009}}

@article{Kedarasetti2022Arterial-vasodi,
	author = {Kedarasetti, R. T. and Drew, P. J. and Costanzo, F.},
	journal = {Fluids and Barriers of the {CNS}},
	note = {{DOI: https://doi.org/10.1186/s12987-022-00326-y; PMID: 35570287; PMCID: PMC9107702}},
	number = {1},
	pages = {34},
	title = {Arterial vasodilation drives convective fluid flow in the brain: a poroelastic model},
	volume = {19},
	year = {2022}}

@article{Kedarasetti2020Functional-hype,
	author = {Kedarasetti, R. T. and Turner, K. L. and Echagarruga, C. and Gluckman, B. J. and Drew, P. J. and Costanzo, F.},
	journal = {Fluids and Barriers of the {CNS}},
	note = {{DOI: 10.1186/s12987-020-00214-3; PMID: 32819402; PMCID: PMC7441569}},
	number = {1},
	pages = {52},
	title = {Functional hyperemia drives fluid exchange in the paravascular space},
	volume = {17},
	year = {2020}}

@Article{Shim2021FiniteElementIm,
  author    = {Shim, Jay J. and Maas, Steve A. and Weiss, Jeffrey A. and Ateshian, Gerard A.},
  journal   = {Journal of Biomechanical Engineering},
  title     = {Finite Element Implementation of Biphasic-Fluid Structure Interactions in Febio},
  year      = {2021},
  issn      = {1528-8951},
  month     = may,
  number    = {9},
  volume    = {143},
  doi       = {10.1115/1.4050646},
  publisher = {ASME International},
}

@Article{Anghileri2005FluidStructureI,
  author   = {Marco Anghileri and Luigi-M.L. Castelletti and Maurizio Tirelli},
  journal  = {International Journal of Impact Engineering},
  title    = {Fluid–Structure Interaction of Water Filled Tanks during the Impact with the Ground},
  year     = {2005},
  issn     = {0734-743X},
  number   = {3},
  pages    = {235--254},
  volume   = {31},
  doi      = {10.1016/j.ijimpeng.2003.12.005},
}

@Article{Huang2026FluidStructureI,
  author    = {Huang, Zheng and Zhao, Zecheng and Wang, Hu and Yang, Yi and Xiong, Wei and Wang, Zhiwen},
  journal   = {Ocean Engineering},
  title     = {Fluid-structure interaction analysis of an underwater flexible compressed air energy storage accumulator based on Arbitrary {L}agrangian-{E}ulerian method},
  year      = {2026},
  issn      = {0029-8018},
  month     = May,
  pages     = {125353},
  volume    = {356},
  doi       = {10.1016/j.oceaneng.2026.125353},
  publisher = {Elsevier BV},
}

@Article{Avalos2024WeakAndStrongSo,
  author    = {Avalos, George and Gurvich, Elena and Webster, Justin T.},
  journal   = {Mathematical Methods in the Applied Sciences},
  title     = {Weak and strong solutions for a fluid‐poroelastic‐structure interaction via a semigroup approach},
  year      = {2024},
  issn      = {1099-1476},
  month     = Oct,
  number    = {4},
  pages     = {4057--4089},
  volume    = {48},
  doi       = {10.1002/mma.10533},
  publisher = {Wiley},
}

@Misc{Guo2024AFullyParalleli,
  author    = {Guo, Shihan and Sun, Yizhong and Wang, Yifan and Yue, Xiaohe and Zheng, Haibiao},
  title     = {A Fully Parallelizable Loosely Coupled Scheme for Fluid-Poroelastic Structure Interaction Problems},
  year      = {2024},
  copyright = {arXiv.org perpetual, non-exclusive license},
  doi       = {10.48550/ARXIV.2409.15618},
  publisher = {arXiv},
}

@Misc{Castro2025WellPosednessOf,
  author    = {de Castro, Amy and Lee, Hyesuk},
  title     = {Well-posedness of a novel {L}agrange multiplier formulation for fluid-poroelastic interaction},
  year      = {2025},
  copyright = {Creative Commons Attribution 4.0 International},
  doi       = {10.48550/ARXIV.2512.08142},
  publisher = {arXiv},
}

@Misc{Cesmelioglu2023AHybridizableDi,
  author    = {Cesmelioglu, Aycil and Lee, Jeonghun J. and Rhebergen, Sander},
  title     = {A hybridizable discontinuous Galerkin method for the coupled {Navier-Stokes/Biot} problem},
  year      = {2023},
  copyright = {arXiv.org perpetual, non-exclusive license},
  doi       = {10.48550/ARXIV.2308.15621},
  publisher = {arXiv},
}

@Article{Badia2009CouplingBiotAnd,
  author    = {Badia, Santiago and Quaini, Annalisa and Quarteroni, Alfio},
  journal   = {Journal of Computational Physics},
  title     = {Coupling {B}iot and {N}avier–{S}tokes equations for modelling fluid–poroelastic media interaction},
  year      = {2009},
  issn      = {0021-9991},
  month     = Nov,
  number    = {21},
  pages     = {7986--8014},
  volume    = {228},
  doi       = {10.1016/j.jcp.2009.07.019},
  publisher = {Elsevier BV},
}

@Article{Bukac2015PartitioningStr,
  author    = {Bukač, M. and Yotov, I. and Zakerzadeh, R. and Zunino, P.},
  journal   = {Computer Methods in Applied Mechanics and Engineering},
  title     = {Partitioning strategies for the interaction of a fluid with a poroelastic material based on a Nitsche’s coupling approach},
  year      = {2015},
  issn      = {0045-7825},
  month     = Aug,
  pages     = {138--170},
  volume    = {292},
  doi       = {10.1016/j.cma.2014.10.047},
  publisher = {Elsevier BV},
}

@Misc{Avalos2025UniquenessOfWea,
  author    = {Avalos, George and Webster, Justin T.},
  title     = {Uniqueness of Weak Solutions for {B}iot-{S}tokes Interactions},
  year      = {2025},
  copyright = {arXiv.org perpetual, non-exclusive license},
  doi       = {10.48550/ARXIV.2502.07061},
  publisher = {arXiv},
}

@Misc{Dalal2024ARobinRobinSpli,
  author    = {Dalal, Aashi and Durst, Rebecca and Quaini, Annalisa and Yotov, Ivan},
  title     = {A {R}obin-{R}obin splitting method for the {S}tokes-{B}iot fluid-poroelastic structure interaction model},
  year      = {2024},
  copyright = {arXiv.org perpetual, non-exclusive license},
  doi       = {10.48550/ARXIV.2409.18910},
  publisher = {arXiv},
}

@Article{Wang2025ALagrangeMultip,
  author    = {Wang, Xing and Yotov, Ivan},
  journal   = {IMA Journal of Numerical Analysis},
  title     = {A {L}agrange multiplier formulation for the fully dynamic {N}avier–{S}tokes–{B}iot system},
  year      = {2025},
  issn      = {1464-3642},
  month     = Sept,
  doi       = {10.1093/imanum/draf074},
  publisher = {Oxford University Press (OUP)},
}

@Article{Hou1989BoundaryConditi,
  author    = {J. S. Hou and M. H. Holmes and W. M. Lai and V. C. Mow},
  journal   = {Journal of Biomechanical Engineering},
  title     = {Boundary Conditions at the Cartilage-Synovial Fluid Interface for Joint Lubrication and Theoretical Verifications},
  year      = {1989},
  month     = feb,
  number    = {1},
  pages     = {78--87},
  volume    = {111},
  doi       = {10.1115/1.3168343},
  publisher = {{ASME} International},
}

@Article{Jannesari2026ModificationOfA,
  author    = {Jannesari, Mohammad and Ghitti, Beatrice and Gluckman, Bruce J. and Costanzo, Francesco},
  journal   = {Biomechanics and Modeling in Mechanobiology},
  title     = {Modification of a poroelastic model for zero porosity: finite element implementation and investigation of fluid mechanics in the perivascular space},
  year      = {2026},
  issn      = {1617-7940},
  month     = jun,
  number    = {3},
  volume    = {25},
  doi       = {10.1007/s10237-026-02070-w},
  publisher = {Springer Science and Business Media LLC},
}

@Article{Costanzo2017AnArbitraryLagr,
  author        = {Costanzo, F. and Miller, S. T.},
  journal       = {Computer Methods in Applied Mechanics and Engineering},
  title         = {An Arbitrary {L}agrangian--{E}ulerian Finite Element Formulation for a Poroelasticity Problem Stemming from Mixture Theory},
  year          = {2017},
  note          = {{DOI: 10.1016/j.cma.2017.05.006}},
  pages         = {64--97},
  volume        = {323},
}

@InBook{Fernandez2009TheDerivationOf,
  author    = {Fernández, Miguel A. and Formaggia, Luca and Gerbeau, Jean-Frédéric and Quarteroni, Alfio},
  pages     = {77–121},
  publisher = {Springer Milan},
  title     = {The derivation of the equations for fluids and structure},
  year      = {2009},
  isbn      = {9788847011526},
  booktitle = {Cardiovascular Mathematics},
  doi       = {10.1007/978-88-470-1152-6_3},
  url       = {http://dx.doi.org/10.1007/978-88-470-1152-6_3},
}

@Article{Beavers1967BoundaryConditi,
  author    = {Beavers, Gordon S. and Joseph, Daniel D.},
  journal   = {Journal of Fluid Mechanics},
  title     = {Boundary conditions at a naturally permeable wall},
  year      = {1967},
  issn      = {1469-7645},
  month     = Oct,
  number    = {1},
  pages     = {197--207},
  volume    = {30},
  doi       = {10.1017/s0022112067001375},
  publisher = {Cambridge University Press (CUP)},
}

@Article{Carraro2013PressureJumpInt,
  author    = {Carraro, T. and Goll, C. and Marciniak-Czochra, A. and Mikelić, A.},
  journal   = {Journal of Fluid Mechanics},
  title     = {Pressure jump interface law for the {S}tokes–{D}arcy coupling: confirmation by direct numerical simulations},
  year      = {2013},
  issn      = {1469-7645},
  month     = Sept,
  pages     = {510--536},
  volume    = {732},
  doi       = {10.1017/jfm.2013.416},
  publisher = {Cambridge University Press (CUP)},
}

@Article{Discacciati2007RobinRobinDomai,
  author    = {Discacciati, Marco and Quarteroni, Alfio and Valli, Alberto},
  journal   = {SIAM Journal on Numerical Analysis},
  title     = {{R}obin–{R}obin Domain Decomposition Methods for the {S}tokes–{D}arcy Coupling},
  year      = {2007},
  issn      = {1095-7170},
  month     = Jan,
  number    = {3},
  pages     = {1246--1268},
  volume    = {45},
  doi       = {10.1137/06065091x},
  publisher = {Society for Industrial & Applied Mathematics (SIAM)},
}

@Article{Saffman1971OnTheBoundaryCo,
  author    = {Saffman, P. G.},
  journal   = {Studies in Applied Mathematics},
  title     = {On the Boundary Condition at the Surface of a Porous Medium},
  year      = {1971},
  issn      = {1467-9590},
  month     = June,
  number    = {2},
  pages     = {93--101},
  volume    = {50},
  doi       = {10.1002/sapm197150293},
  publisher = {Wiley},
}

@Article{Jones1973LowReynoldsNumb,
  author    = {Jones, I. P.},
  journal   = {Mathematical Proceedings of the Cambridge Philosophical Society},
  title     = {Low Reynolds number flow past a porous spherical shell},
  year      = {1973},
  issn      = {1469-8064},
  month     = Jan,
  number    = {1},
  pages     = {231--238},
  volume    = {73},
  doi       = {10.1017/s0305004100047642},
  publisher = {Cambridge University Press (CUP)},
}

@InCollection{Showalter2005PoroelasticFilt,
  author    = {Showalter, R. E.},
  booktitle = {Control Theory of Partial Differential Equations},
  publisher = {Chapman \& Hall/CRC},
  title     = {Poroelastic Filtration Coupled to {S}tokes Flow},
  year      = {2005},
  address   = {Boca Raton, FL},
  pages     = {229--241},
  series    = {Lecture Notes in Pure and Applied Mathematics},
  volume    = {242},
  doi       = {10.1201/9781420028317.ch16},
}

@Article{Ateshian2010FiniteElementAl,
  author    = {Ateshian, Gerard A. and Maas, Steve and Weiss, Jeffrey A.},
  journal   = {Journal of Biomechanical Engineering},
  title     = {Finite Element Algorithm for Frictionless Contact of Porous Permeable Media Under Finite Deformation and Sliding},
  year      = {2010},
  issn      = {1528-8951},
  month     = apr,
  number    = {6},
  volume    = {132},
  doi       = {10.1115/1.4001034},
  publisher = {ASME International},
}

@Article{Shim2021AHybridBiphasic,
  author    = {Jay J. Shim and Gerard A. Ateshian},
  journal   = {Archive of Applied Mechanics},
  title     = {A Hybrid Biphasic Mixture Formulation for Modeling Dynamics in Porous Deformable Biological Tissues},
  year      = {2021},
  month     = jan,
  number    = {2},
  pages     = {491--511},
  volume    = {92},
  doi       = {10.1007/s00419-020-01851-8},
  publisher = {Springer Science and Business Media {LLC}},
}

@article{Wick2011Fluid-Structure-0,
	author = {Wick, T.},
	doi = {10.1016/j.compstruc.2011.02.019},
	journal = {Computers and Structures},
	note = {{DOI: 10.1016/j.compstruc.2011.02.019}},
	number = {13--14},
	pages = {1456--1467},
	title = {Fluid-Structure Interactions using Different Mesh Motion Techniques},
	volume = {89},
	year = {2011},
}

@techreport{Salari2000Code-Verification-0,
	author = {Salari, K. and Knupp, P.},
	institution = {Sandia National Laboratories},
	title = {Code Verification by The Method of Manufactured Solutions},
	type = {SAND2000-1444},
	year = {2000}}
\bibliographystyle{plain}
\end{document}